\documentclass[aps,prb,twocolumn,superscriptaddress,amsmath,amssymb,floatfix]{revtex4}
\usepackage{graphicx}
\usepackage{dcolumn}
\usepackage{bm}

\newcommand{\pdag}{{\phantom{\dagger}}}
\newcommand{\ybco}{YBa$_2$Cu$_3$O$_{y}$}
\newcommand{\ndlsco}{La$_{1.6-x}$Nd$_{0.4}$Sr$_x$CuO$_{4}$}
\newcommand{\eulsco}{La$_{1.8-x}$Eu$_{0.2}$Sr$_x$CuO$_{4}$}
\newcommand{\lsco}{La$_{2-x}$Sr$_x$CuO$_{4}$}
\newcommand{\lbco}{La$_{2-x}$Ba$_x$CuO$_{4}$}
\newcommand{\lbcoo}{La$_{15/8}$Ba$_{1/8}$CuO$_4$}
\newcommand{\bsco}{Bi$_2$Sr$_{2}$CaCu$_2$O$_{8+\delta}$}
\newcommand{\cnco}{Ca$_{2-x}$Na$_{x}$CuO$_{2}$Cl$_2$}

\newcommand{\Tc}{T_{\rm c}}
\newcommand{\Tch}{T_{\rm ch}}
\newcommand{\Tsp}{T_{\rm sp}}
\newcommand{\bk}{{\bf k}}

\begin{document}

\title{Quasiparticle Nernst effect in stripe-ordered cuprates}

\author{Andreas Hackl}
\email[]{ah@thp.uni-koeln.de}
\affiliation{Institut f\"ur Theoretische Physik, Universit\"at zu K\"oln,
Z\"ulpicher Stra{\ss}e 77, 50937 K\"oln, Germany}
\author{Matthias Vojta}
\affiliation{Institut f\"ur Theoretische Physik, Universit\"at zu K\"oln,
Z\"ulpicher Stra{\ss}e 77, 50937 K\"oln, Germany}
\author{Subir Sachdev}
\affiliation{Department of Physics, Harvard University, Cambridge, Massachusetts 02138, USA}

\date{\today}


\begin{abstract}
Experiments on underdoped cuprate superconductors suggest an intricate relation
between the normal-state Nernst effect and stripe order:
The Nernst signal appears enhanced near 1/8 hole doping and its onset temperature
scales with the stripe-ordering temperature over some range of doping.
Here, we employ a phenomenological quasiparticle model to calculate the normal-state
Nernst signal in the presence of stripe order.
We find that Fermi pockets caused by translational symmetry breaking lead to a strongly
enhanced Nernst signal, with a sign depending on the modulation period of the ordered
state and other details of the Fermi surface.
This implies differences between antiferromagnetic and charge-only stripes.
We also analyze the anisotropy of the Nernst signal and
compare our findings with recent data from \ndlsco\ and \ybco.
\end{abstract}

\maketitle


\section{\label{intro} Introduction}

The pseudogap regime of cuprate superconductors\cite{timusk}
has remained mysterious despite more than
two decades of intense research.
Among the various proposed explanations for the observed
suppression of spectral weight below the doping-dependent pseudogap temperature $T^\ast$
are phase-incoherent Cooper pairing, symmetry-breaking orders competing with
superconductivity, exotic fractionalized states, and short-range singlet correlations
as precursor to the half-filled Mott insulator.\cite{norman_rev,leermp}

Nernst effect measurements have been established as an interesting probe of pseudogap
physics. The Nernst signal, measuring the transverse voltage induced by a thermal gradient,
is typically small in conventional metals.
Large positive Nernst signals are known to arise from the motion of vortices in type-II
superconductors.\cite{maki,vortex}
In underdoped cuprates, with experiments performed on a variety of different families,
the Nernst signal has been found to rise upon cooling, with an
onset temperature significantly above the superconducting $\Tc$
(although it is difficult to define a sharp onset).\cite{xu00,wang06}
The data have been commonly interpreted as evidence for fluctuating Cooper pairs above $\Tc$;
this interpretation appears supported by the observation of fluctuating diamagnetism
which often varies in proportion to the Nernst coefficient.\cite{nernst_diamag}
As function of doping, the onset temperature of the Nernst signal is maximum around
10\%--15\% doping and appears to lie below the $T^\ast$ line identified by other probes,
in particular for doping $x<10\%$. A plausible conclusion is that fluctuating Cooper
pairs do not account for all of the cuprate pseudogap.
On the theory side, the Nernst signal arising from Gaussian (i.e. amplitude) pairing fluctuations
has been calculated,\cite{huse02,ussi03} and theoretical treatments of short-lived vortex
(i.e. phase) fluctuations have been put forward as well.\cite{podol07,raghu08}
Meanwhile it has also been established that a large Nernst signal can occur in
metals with a small Fermi energy, in particular in the presence of electron and hole
pockets.\cite{nernst_fl} In underdoped cuprates, this situation has
been discussed especially in a scenario of $d$-density wave order.\cite{oga04,tewari09,zhang09}

Recently, a more detailed investigation\cite{taill09} of the Nernst effect in the \lsco\
(or ``214'') family revealed a new piece of information:
In \ndlsco, which is known to display static stripe
order below a temperature $\Tch$, an additional (positive) peak or shoulder in the temperature
dependence of the Nernst signal could be identified, located at an elevated temperature
and distinct from the low-temperature signal ascribed to superconducting fluctuations.
As this high-temperature feature appears to follow the charge-ordering temperature $\Tch$
upon variation of the doping level, it has been attributed to a Fermi-surface
reconstruction due to density-wave order.

A Fermi surface reconstruction due to density-wave order also appears as a candidate
explanation for quantum oscillations, observed in large fields on underdoped \ybco\
samples.\cite{doiron07,sebastian08,yelland08}
Indeed, neutron-scattering experiments indicate field-induced incommensurate spin-density
wave order in this cuprate family.\cite{haug_mf}
On the theory side, concrete symmetry-breaking patterns have been proposed to explain the
observed quantum oscillations.\cite{millis07,harrison09,sudip}
Among the various ordering phenomena, stripe order plays a prominent role:
While first established in certain 214 cuprates and initially considered to be special
to this family, signatures of (possibly fluctuating or disordered) stripes have meanwhile
been found in a variety of cuprates over a significant doping range.\cite{stripe_rev1,stripe_rev2}\\
Recent transport phenomenology in \ybco\ at hole concentration $x=0.12$ has
shown close similarities to the 214 cuprates, including a sign change of
the Seebeck coefficient at $T \simeq 50$\,K and a strongly enhanced normal-state
Nernst signal accessed by strong magnetic fields of up to $28$\,T. \cite{daou2009}
Interestingly, the sign of the normal-state Nernst effect is negative even at
lowest temperatures, and theoretical explanations of how the sign of
the Nernst coefficent is related to Fermi-surface reconstruction are lacking.

Taken together, these developments suggest that density-wave order plays a vital role in
the phenomenology of underdoped cuprates. It is thus of timely importance to clarify which
experimental results can be understood in terms of density-wave order of conventional
quasiparticles, and where physics beyond the quasiparticle picture needs to be invoked.
In this paper we shall present a theoretical calculation of the normal-state
Nernst signal in the presence of uni-directional spin and charge density wave
(i.e. stripe) order.

In fact, in Ref.~\onlinecite{blackhole}, it was argued that charge-density-wave fluctuations were
important for the pseudogap Nernst signal, and a general hydrodynamic discussion
was presented at moderately high temperatures above a charge-ordering critical point.
However, a specific comparison with experiment requires that we go to lower temperatures
and consider the coherent dynamics of electronic quasiparticles. Such an analysis was provided
for the electron-doped cuprates in Ref.~\onlinecite{hackl}, where it gave a good
account for the experimental observations.\cite{Greene2007}

It is the purpose of the present paper to apply such a quasiparticle analysis to the
hole-doped case, by combining a mean-field description of stripe order with a Boltzmann approach
to transport.
At low temperatures, the normal-state Nernst signal varies linearly with $T$,
and we shall discuss the sign and magnitude of this piece in connection with the Fermi-surface
pockets induced by the density-wave order. The focus will be on order with
real-space periods 4 (8) and 8 (16) in the charge (spin) sector,
being appropriate for 1/8-doped \lsco\ and \ybco\ close to $y=6.5$,
respectively.
In the light of a recent experiment\cite{taill10b} which examined the spatial
anisotropy of the Nernst coefficient in \ybco\ we shall calculate this quantity
for thermal gradients both perpendicular and parallel to the stripes.

We note that recent papers have provided a detailed discussion of the effect of
stripe order on quantum oscillations\cite{millis07} and the Hall effect,\cite{linmillis}
using mean-field models similar to ours below.
For both observables, reasonable agreement with experiment was
pointed out, and we refer the reader to those papers for details.
Below, we shall make use of the results of Refs.~\onlinecite{millis07,linmillis}
when appropriate.

\subsection{Outline}

The remainder of this paper is organized as follows: In Sec.~\ref{model} we describe the
microscopic mean-field model for stripe order and discuss the Boltzmann transport
formalism which we shall use to evaluate the low-temperature Nernst effect.
Sec.~\ref{nernst} contains our main results for the Fermi-surface reconstruction and the
Nernst signal in stripe phases with a real-space period of eight sites in the spin
sector, this includes the doping level of 1/8 where stripes are particularly stable. We
shall discuss the effect of modulations in the spin and charge sectors separately, and
also distinguish between site-centered and bond-centered stripes.
These considerations will be extended to doping below 1/8 in Sec.~\ref{underdoped}, where
the real-space modulation period is larger.
A summary and comparison to experimental data is given in Sec.~\ref{experiments}.


\section{\label{model} Model and Formalism}

To calculate the normal-state quasiparticle Nernst effect,
we consider electrons moving on a square lattice of unit lattice constant,
with the two-dimensional dispersion given by
\begin{eqnarray}
\varepsilon_{ \bf k} = &-& 2 t_1 (\cos k_x +\cos k_y ) - 4 t_2 \cos k_x \cos k_y \nonumber\\
                &-&2 t_3 (\cos 2 k_x +\cos 2 k_y) \ .
\label{dispersion}
\end{eqnarray}
For all numerical calculations, we will use the parameters $t_1=0.38$\,eV, $t_2=-0.32 t_1$
and $t_3=-0.5 t_2$,\cite{norman94,Andersen1995} chosen to reproduce the Fermi surface measured in
photo\-emission experiments.
The two-dimensional electron density is $n=1-x$ per unit cell.
We shall assume a quasiparticle description with the dispersion \eqref{dispersion}
to be a reasonable approximation in the regimes of interest, i.e.,
either at low temperatures and strong fields or above the superconducting $\Tc$ at
smaller fields.
Effects of pseudogap physics beyond quasiparticles, like phase-fluctuating Cooper pairs,
will be briefly discussed in Sec.~\ref{pairing}.

\subsection{Stripe order}
\label{stripes}

The term ``stripe'' shall be used synonymously for uni-directional spin-density and
charge-density wave order.
A spin density wave (SDW) is specified by a vector order parameter
$\phi_{s\alpha}({\bf r},\tau)$, $\alpha=x,y,z$, and the spin density modulation is given by
\begin{equation}
\langle S_\alpha ( {\bf R},\tau ) \rangle = Re \bigl[ e^{i{\bf Q}_s\cdot {\bf R}} \phi_{s\alpha} ({\bf R},\tau ) \bigr]
\label{sdw}
\end{equation}
with ordering wavevector ${\bf Q}_s$.
As charge density wave (CDW) we will denote a state with modulations in observables $\rho$
which are invariant under spin rotation and time reversal, such as
site or bond charge density, kinetic energy, or pairing amplitude.
A CDW is described by a scalar order parameter $\phi_c({\bf r},\tau)$, such that
\begin{equation}
\langle\rho({\bf R},\tau ) \rangle = \rho_0 + Re \bigl[ e^{i{\bf Q}_c\cdot {\bf R}} \phi_c ({\bf R},\tau ) \bigr]
\label{cdw}
\end{equation}
where $\rho_0$ is the background density.
If the SDW order in Eq. (\ref{sdw}) is collinear, it has an associated spin-singlet order parameter,
i.e. it induces a CDW with wavevector ${\bf Q}_c=2{\bf Q}_s$.\cite{kivelson1998}

Historically, incommensurate SDW order in cuprates was first found\cite{jt95}
in neutron scattering experiments on \ndlsco,
with wavevectors ${\bf Q}_{sx} = 2\pi(0.5\pm\epsilon_s,0.5)$ and ${\bf Q}_{sy} = 2\pi(0.5,0.5\pm\epsilon_s)$.
Corresponding charge order at ${\bf Q}_{cx} = 2\pi(\pm\epsilon_c,0)$ and ${\bf Q}_{cy} =
2\pi(0,\pm\epsilon_c)$, with $\epsilon_s=2\epsilon_c$, was detected as well.
Subsequently, such stripe order, with co-existing SDW and CDW,
was also established to exist in \eulsco\ and \lbco.
Whereas in \lbco\ the order is confined to a narrow doping range around $x=1/8$,
it appears to extend from low doping up to 20\% in \ndlsco\ and \eulsco.\cite{stripe_rev2}
In \lsco\ with $x<0.13$ and in YBa$_2$Cu$_3$O$_{6.35}$,
quasi-static SDW order was found,\cite{waki99,buyers06,haug_635}
while for larger doping incommensurate dynamic spin fluctuations
exist.\cite{cheong91,stock04,hinkov08}
In both cases, strong magnetic fields applied to superconducting samples
can enhance and even induce SDW order,\cite{lake01,chang09,haug_mf}
suggesting a competition between SDW and superconducting orders.
Static order in the charge sector has not been detected in \lsco,
while reports on charge order in \ybco\ remained controversial.\cite{mook02,buyers06}
Using scanning tunneling microscopy (STM) techniques,
static short-range bond-centered modulations in the charge sector have been detected
on the surface of \bsco\ and \cnco.\cite{kapi03a,kohsaka07}
The modulation period was close to four lattice spacings, similar to the charge order in
striped 214 compounds with doping $x\geq 1/8$.
The STM data appear to be well described by modulations
in the kinetic-energy terms,\cite{podolsky2003} which moreover appear to have a large $d$-wave
component.\cite{vojta2008}
Note that in both \bsco\ and \cnco\ the charge order appears to exist without long-range
magnetic order, although spin-glass-like magnetism has been reported in \cnco.\cite{ohishi05}

With regard to Nernst effect and quantum oscillation measurements,
we may expect modulations in the spin sector to be important for the
Fermi-surface reconstruction, as SDW order occurs in both 214 and \ybco\ compounds in
strong fields.

\subsection{Mean-field theory}
\label{mf}

The ordered states shall be described in a mean-field picture, where quasiparticles with
the dispersion~\eqref{dispersion} are subject to a periodic modulation in the site
chemical potential or bond kinetic energy.
Philosophically, we assume that both the quasiparticles and the modulation arise
from a microscopic Hubbard or $t$--$J$ model at intermediate or strong coupling.
Suitable self-consistent mean-field calculations
have been reported in the literature, with results which appear broadly consistent with the
experimental phenomenology (for a review, see e.g. Ref.~\onlinecite{stripe_rev2}).
Here, we find it appropriate to combine this previous knowledge with experimental input
(e.g. on the wavevector and magnitude of modulations), and hence we will add the periodic
modulations to the quasiparticle Hamiltonian ``by hand'', i.e.,
without performing a self-consistent evaluation.
We note that SDW order can in principle be obtained in a controlled manner at weak
coupling,\cite{rice70,fawcett1988,bazaliy04} whereas CDW order in cuprates is likely
a strong-coupling phenomenon, with additional stabilization by lattice degrees of
freedom.\cite{stripe_rev1,stripe_rev2}

\begin{figure}
\includegraphics[width=7.0cm]{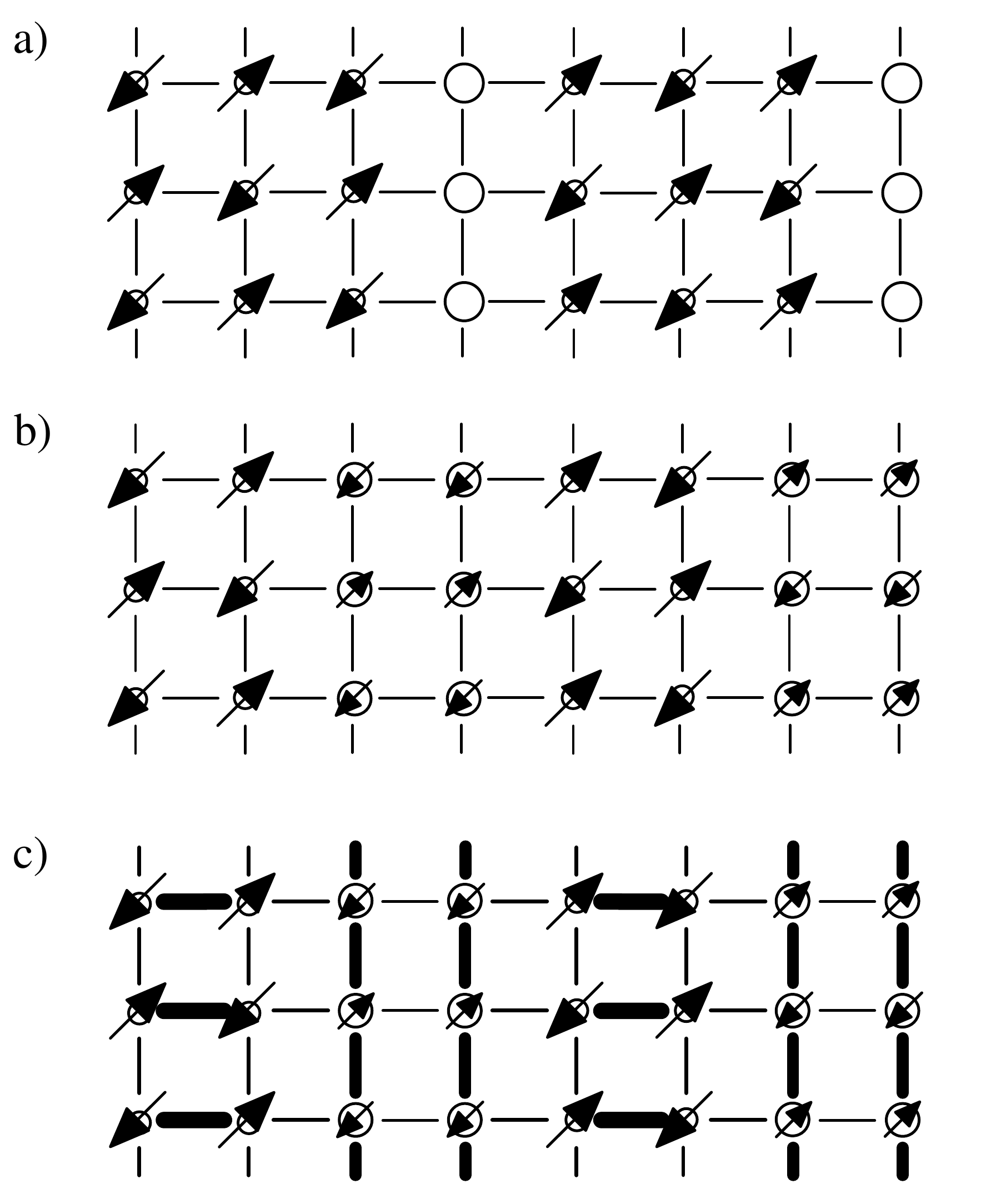}
\caption{
\label{antiphase}
Real-space structure of a) site-centered and b),c) bond-centered stripes
with period-4 (period-8) order in the charge (spin) sector.
Shown are spin and charge distributions, with the circle radii corresponding
to on-site hole densities.
In panel c), showing ``valence-bond'' stripes,\cite{vojta2008,vojta2008b}
the structure of spin-singlet bond modulations is shown as well which
has a dominant $d$-wave form factor.
}
\end{figure}

In the spin sector, we shall restrict our attention to collinear order.
Such order leads to a scattering potential $V_s$ that connects a quasiparticle
with momentum ${\bf k}$ with all quasiparticle momenta ${\bf k}\pm n {\bf Q}$ for integer $n$.
(The same applies to charge order with wavevector ${\bf Q}_c$ and a scattering potential
$V_c$.)
As has been discussed for Cr, the Fermi surface reconstruction
due to collinear SDW order is caused by a hierarchy of gaps of order $2\Delta_m \sim 2V_s^m/t^{m-1}$
opening at the crossing points of bands $\varepsilon_{\bk+n{\bf Q}}$ and
$\varepsilon_{\bk+(n\pm m){\bf Q}}$, where $V_s$ is the amplitude of
the spin potential.\cite{fawcett1988}
As long as $V_s,V_c \ll t$, the Fermi surface is
well described by including the lowest-order gap only, and we will neglect
all matrix elements with $m>1$ in the scattering potentials $V_c$ and $V_s$.
In the mean-field Hamiltonian, we shall use the following terms
describing the density waves; cartoons pictures of the resulting stripe order are shown
in Fig.~\ref{antiphase}.

{\it Charge density wave. \/}
A CDW is described by
\begin{equation}
\hat{V}_1= \sum_{{\bf k},\sigma} \bigl( V_c({\bf k}) c_{{\bf k}+{\bf Q}_c\sigma}^\dagger c_{{\bf k}\sigma}^\pdag
+h.c. \bigr)
\label{vcdw}
\end{equation}
where $V_c({\bf k})$ is in general complex.
For the site-centered case, we modulate the on-site (Hartree-Fock) chemical potentials
such that maxima/minima are located on lattice sites, i.e., with a real $V_c({\bf k}) \equiv -V_c$.
A bond-centered CDW with on-site modulations is characterized by $V_c( \bk) \equiv -V_c
e^{-i Q_c/2}$; for modulations in the kinetic energy with primarily $d$-wave form factor
we have $V_c( \bk) = -\delta t (\cos(k_x+\frac{Q_c}{2})-\cos k_y ) e^{-iQ_c/2}$;
in both cases ${\bf Q}_c = (Q_c,0)$.
In the following, we shall primarily consider the latter $d$-wave bond modulations,
which arise in a scenario of valence-bond solid formation\cite{vs99,mv02} and
have been argued\cite{vojta2008}
to be consistent with the STM data of Ref.~\onlinecite{kohsaka07}.

{\it Collinear spin density wave. \/}
Choosing the spin quantization axis in $z$ direction, we have in general
\begin{equation}
\hat{V}_2= \sum_{\bk,\sigma} \sigma \bigl( V_s(\bk) c_{\bk+{\bf Q}_s \sigma}^\dagger c_{{\bf k}\sigma}^\pdag
+h.c. \bigr).
\label{vsdw}
\end{equation}
A site-centered SDW has again a real $V_s({\bf k}) \equiv V_s$,
whereas a bond-centered SDW is captured by $V_s(\bk) \equiv -V_s
(1 + e^{-iQ_c/2})/(2 \cos(Q_c/4))$ where ${\bf Q}_s = (\pi\pm Q_c/2,\pi)$.
The complex phases of the mean fields in Eqs.~\eqref{vcdw} and~\eqref{vsdw}
have been chosen such that the resulting order parameters $\phi_c$ and $\phi_s^2$
are in-phase. Moreover, with positive $V_c$ (site-centered) and positive $\delta t$
(bond-centered) the resulting modulations are such that the electron density is
small where the magnitude of the magnetic moment is small (i.e. near the anti-phase
domain walls),\cite{millisfoot}
as in Fig.~\ref{antiphase}.

As the uni-directional density waves break the $90^{\circ}$ rotation symmetry of
the underlying square lattice, the $s$-wave and $d_{x^2-y^2}$ representations of the
point group mix.
This implies that the solution of a mean-field Hamiltonian with modulated on-site potentials
(which may be dubbed $s$-wave) will also contain symmetry-compatible modulations on the bonds,
with inequivalent horizontal and vertical bonds (i.e. a $d$-wave component).
Vice versa, the solution of a mean-field Hamiltonian with $d$-wave bond modulations will display
a finite on-site charge density modulation.
Also, solving a Hamiltonian with a collinear SDW modulation only will lead to a
CDW with ${\bf Q}_c=2{\bf Q}_s$.

In Secs.~\ref{nernst} and~\ref{underdoped} we shall present results separately for the
cases of spin-only, charge-only, and combined spin and charge modulations in the
mean-field Hamiltonian. While the charge-only case corresponds to a situation without
broken spin symmetry, the spin-only and the combined spin and charge cases have the same
symmetry, but the former is to be understood as density-wave order {\em driven} by the spin sector,
with charge order being parasitic.

\subsection{Semiclassical transport}
\label{transp}

The Nernst effect is measured as a transverse electrical response to a thermal gradient,
which can also generate a longitudinal electrical voltage known as thermopower.
In experiment, the electric field can be applied by allowing for
a weak spatial dependence in the chemical potential $\mu$
(which is then, formally, the electrochemical potential) with $2e{\bf E}=-\nabla \mu$,
while the temperature gradient describes a similar weak spatial dependence in
$T$. The interplay of electrical and thermal effects necessarily implies three conductivity tensors $\hat{\sigma}$,
$\hat{\alpha}$ and $\hat{\kappa}$, which relate charge current ${\bf J}$
and heat current ${\bf Q}$ to electric field, ${\bf E}$ and thermal gradient,
${\bf \nabla }T$ vectors:
\begin{equation}
\left(
\begin{array}{c}
{\bf J} \\
{\bf Q} \\
\end{array}
\right)
=
\left(
\begin{array}{cc}
\hat{\sigma}  & \hat{\alpha} \\
T\hat{\alpha} &  \hat{\kappa}\\
\end{array}
\right)
\left(
\begin{array}{c}
{\bf E} \\
-{\bf \nabla} T\\
\end{array}
\right)
\label{thermoelectrics}
\end{equation}
It is the off-diagonal component $\hat{\alpha}$ which relates electrical currents
and fields to thermal currents and gradients. To measure this quantity,
appropriate boundary conditions for the currents and applied fields have to be
obeyed. The Nernst response is defined as the electrical field induced
by a thermal gradient in the absence of an electrical current, and is given in
linear response by the relation ${\bf E}= -\hat{\vartheta } {\bf \nabla} T$.
In absence of charge current (i.e. when ${\bf J}=0$), Eq. (\ref{thermoelectrics})
yields:
\begin{equation}
{\bf E}=\hat{\sigma}^{-1} \hat{\alpha} {\bf \nabla} T \ .
\end{equation}
Therefore, the Nernst signal defined as the transverse voltage $E_y$ generated by a thermal gradient $\nabla_x T$
reads
\begin{equation}
\vartheta_{yx}=-\frac{\sigma_{xx}\alpha_{yx}-\sigma_{yx}\alpha_{xx}}{\sigma_{xx}\sigma_{yy}-\sigma_{xy}\sigma_{yx}}
\label{nernstsignal}
\end{equation}
and $\vartheta_{xy}$ is obtained from $x \leftrightarrow y$. For a magnetic field
$\vec{B}=B \hat{z}$ in $z$ direction, the Nernst { \it coefficient \/} is usually defined as $\nu_{yx}=\vartheta_{yx}/B$,
which tends to become field-independent at small $B$.
We employ a sign convention such that the vortex Nernst coefficient is always positive.
This is achieved by the experimentally used convention that the three vectors ${\bf E}$, $\nabla T$ and ${\bf B}$
form a right-handed system for the measurements of both $\nu_{xy}$ and $\nu_{yx}$.
In general, the Nernst signal can be negative or positive,
for example if it is caused by the flow of charged quasiparticles.

We assume that the low-temperature DC transport can be described by the Boltzmann equation
in relaxation-time approximation\cite{Ziman}
\begin{eqnarray}
&&\biggl[-\frac{e}{\hbar c} ({\bf v}_\bk\times {\bf B}) \cdot \nabla_\bk +\frac{1}{\tau_\bk} \biggr] g_{\bk}= \nonumber\\
&&\biggl[ -e {\bf v}_\bk {\bf E} -( \varepsilon_\bk -\mu ) {\bf v}_\bk \frac{\nabla_{\bf r} T}{T}\biggr]\biggl( -\frac{\partial f_\bk^{0}}{\partial \varepsilon_\bk}\biggr) \ .
\label{boltzmann1}
\end{eqnarray}
The right-hand side has been linearized in both temperature gradient and electric field,
assuming that those are weak and spatially uniform.
The solution of Eq.~\eqref{boltzmann1} is the deviation $g({\bf k})$ of the non-equilibrium
distribution function $f({\bf k})$
from the equilibrium Fermi distribution $f_0({\bf k})=(1+\exp[\beta (\varepsilon_\bk-\mu)])^{-1}$.
We further assume, as is appropriate for low temperatures, that the relaxation is mainly due
to randomly distributed impurities with a low density,\cite{impfoot}
leading to a constant relaxation time $\tau_{\bf k}\equiv \tau_0$.
This approximation is known to fail in presence of antiferromagnetic
fluctuations, which lead to interaction induced drag between quasiparticles.\cite{kontani2006}
Therefore, the assumption of a single-particle relaxation rate is
restricted to temperatures below the ordering temperatures of spin and charge order.

From Eq.~\eqref{boltzmann1}, the non-equilibrium distribution function $g(\bk)$ is now readily obtained as
\begin{equation}
g_\bk=A_\bk^{-1} \biggl[- e {\bf v}_\bk {\bf E} -(\varepsilon_\bk-\mu) {\bf v}_\bk \frac{\nabla_{\bf r} T}{T}\biggr] \biggl( -\frac{\partial f_\bk^{0}}{\partial \varepsilon_\bk}\biggr)
\label{formalsolution}
\end{equation}
where the operator
\begin{equation}
A_\bk=\biggl[ -\frac{e}{\hbar c} ({\bf v}_\bk\times {\bf B}) \cdot \nabla_\bk +\frac{1}{\tau_{\bk}} \biggr]
\end{equation}
has been defined. From the solution~\eqref{formalsolution}, the electrical and thermal
currents ${\bf J}$ and ${\bf Q}$ can be calculated as
\begin{eqnarray}
{\bf J}&=&-e\sum_\bk {\bf v}_\bk g_\bk \ , \nonumber\\
{\bf Q}&=&\sum_\bk {\bf v}_\bk (\epsilon_\bk-\mu ) g_\bk \ .
\end{eqnarray}
According to Eq.~\eqref{thermoelectrics}, the transport tensors are determined from
\begin{eqnarray}
\sigma_{\mu \nu} &=& 2e^2 \sum_\bk v_\bk^\mu A_\bk^{-1} v_\bk^\nu \biggl( -\frac{\partial f_\bk^{0}}{\partial \varepsilon_k}\biggr)\nonumber\\
\alpha_{\mu \nu} &=& -\frac{2 e}{T} \sum_\bk v_\bk^\mu (\varepsilon_\bk -\mu ) A_\bk^{-1} {\bf v}_\bk^\nu \biggl( -\frac{\partial f_\bk^{0}}{\partial \varepsilon_\bk}\biggr) \ .
\label{transportcoeff}
\end{eqnarray}
In the usual manner, $A_\bk^{-1}$ can be arranged as a perturbative expansion in the magnetic field ${\bf B}$
(Ref.~\onlinecite{Ziman})
in order to obtain transport coefficients that do not depend on ${\bf B}$. For this purpose we
define $A_\bk=K_\bk+M_\bk^B$ where $K_\bk=\tau_\bk^{-1}$ and $M_\bk^B$ the rest.
Then
\begin{equation}
A_\bk^{-1} =K_\bk^{-1} -K_\bk^{-1} M^B_\bk K_\bk^{-1} + \mathcal{O}(B^2) \ .
\label{zenerjones}
\end{equation}
The diagonal entries in Eq.~\eqref{transportcoeff} are obtained from the zeroth order in $B$ in Eq.~\eqref{zenerjones},
while the lowest-order contribution to the off-diagonal coefficients arises from the linear order in $B$
in the expansion~\eqref{zenerjones}.
To this accuracy, the expressions~\eqref{transportcoeff} can be simplified in form of the expressions
\begin{eqnarray}
\alpha_{xx} &=&  \frac{2e}{T} \sum_{{\bf k}, n} \frac{\partial f_{\bf k}^0}{\partial \varepsilon_n({\bf k})}
\varepsilon_n({\bf k})  \tau_0 (v_{\bf k}^x)^2    \nonumber\\
\alpha_{xy} &=&  \frac{2e^2B}{T\hbar c} \sum_{{\bf k}, n} \frac{\partial f_{\bf k}^0}{\partial \varepsilon_n({\bf k})}
\varepsilon_n({\bf k}) \tau_0^2 v_{\bf k}^x
\biggl[ v_{\bf k}^y \frac{\partial v_{\bf k}^y}{\partial k_x} -  v_{\bf k}^x \frac{\partial v_{\bf k}^y}{\partial k_y}\biggr]\nonumber\\
\sigma_{xx} &=&  -2e^2 \sum_{{\bf k}, n} \frac{\partial f_{\bf k}^0}{\partial \varepsilon_n({\bf k})}
\tau_0 (v_{\bf k}^x)^2  \nonumber\\
\sigma_{xy} &=& -2 \frac{e^3 B}{\hbar c} \sum_{{\bf k}, n} \frac{\partial f_{\bf k}^0}{\partial \varepsilon_n({\bf k})}
\tau_0^2 v_{\bf k}^x  \biggl[ v_{\bf k}^y \frac{\partial v_{\bf k}^y}{\partial k_x} -  v_{\bf k}^x \frac{\partial v_{\bf k}^y}{\partial k_y}\biggr]
\label{boltzmann}
\end{eqnarray}
which is the result we employ in the rest of the paper. On general grounds, the Hall conductivities obey $\sigma_{xy}=-\sigma_{yx}$.
Such a relation does not hold for $\alpha_{xy,yx}$ in general. In the low-$T$ limit of the Boltzmann
Eq.~\eqref{boltzmann}, however, $\alpha_{xy}=-\alpha_{yx}$ follows from Eq.~\eqref{mottrelation}.

It is important to note that the transport quantities in Eq. (\ref{boltzmann}) describe
transport within a single layer of a cuprate sample only. Apart from weak interlayer
coupling (which we shall neglect here), the most important aspect of multiple layers is
in the stripe directions.
In 214 cuprates with a LTT lattice structure, like \ndlsco,
the stripe orientation of neighboring layers is believed to follow the low-temperature
tetragonal (LTT) distortion in-plane pattern and hence alternates from layer to layer.
Thus, transport quantities have to be averaged over neighboring layers,
in order to obtain the correct bulk transport coefficients.
In contrast, rotation symmetry breaking in \ybco\ compounds can be expected to
have the same orientation in all layers, due to the presence of CuO chains in this
material. Consequently, a single-layer description of transport is sufficient.
In the following, we shall discuss both the single-layer Nernst coefficients
$\nu_{yx,xy}$ as well as a symmetrized version $\nu=(\nu_{xy}+\nu_{yx})/2$
obtained from averaging over layers.

Let us make a few more remarks on the validity of the transport equations
(\ref{boltzmann}); a more extensive discussion can be found in Ref.~\onlinecite{hackl}.
By neglecting the energy dependence of the relaxation time,\cite{impfoot}
one neglects contributions to the Nernst signal which are proportional to the energy derivative of the
relaxation time, defined by the derivative with respect to the position of the Fermi surface,
$\partial \tau / \partial \mu |_{E_F}$.
This can be seen from the Mott relation
\begin{equation}
\alpha_{ij}=-\frac{\pi^2}{3} \frac{k_B^2T}{e} \frac{\partial \sigma_{ij}}{\partial \mu} |_{E_F} \ ,
\label{mottrelation}
\end{equation}
which is valid at temperatures sufficiently below the Fermi temperature. By employing the Mott relation
in equation (\ref{nernstsignal}), one can see that a sizeable contribution to the Nernst signal
from an energy dependence of the relaxation time requires that $\sigma_{xx}\alpha_{yx}$ and $\sigma_{yx}\alpha_{xx}$
have the same order of magnitude. From experiments on the hole-doped cuprates, it is known that the contribution
of $-\alpha_{yx}/\sigma_{yy}$ is dominating the low-temperature Nernst signal in
order of magnitude, \cite{wang01} although this signal is dominated by the vortex contribution.
In the electron-doped cuprates, magnetic fields can suppress the vortex contribution to
the Nernst signal with a Nernst signal that remains dominated by
the contribution of $\sigma_{yx}\alpha_{xx}/(\sigma_{xx}\sigma_{yy})$,~\cite{Greene2007}
and it appears reasonable to neglect an energy dependence of the relaxation time.
In addition, various contributions of interband transitions to quasiparticle transport are neglected in the
transport equations  (\ref{boltzmann}). These can result from thermal excitations, magnetic breakdown
or also scattering on impurities. We will discuss corrections due to these effects where necessary.
In general, such effects are small in the experimentally relevant regimes
as long as stripe order induces band gaps of order $0.1$\,eV.

In order to integrate the transport equations~\eqref{boltzmann} we calculated the
first-order and second-order partial derivatives of the eigenvalues for each ${\bf k}$-point of the
reduced Brillouin zone by an iterative procedure~\cite{andrew} and discretized the
Brillouin zone integrals with a mesh around the Fermi surface of an energy width
proportional to temperature and extrapolated the result to zero temperature. In this
limit, it follows from Eq.~\eqref{mottrelation} that the Nernst signal becomes linear in
temperature, with a prefactor controlling sign and magnitude of the Nernst
signal.
The relaxation rate $\tau_0^{-1}$ remains a parameter in this low-temperature calculation,
with the Nernst signal being proportional to $\tau_0$.
Below we shall briefly discuss the temperature dependence of the Nernst signal as well;
there we will employ suitable phenomenological parametrizations of $\tau(T)$.


\section{ Nernst effect from stripe order for $x\geq 1/8$}
\label{nernst}

As discussed in Sec.~\ref{stripes}, for 214 cuprates with doping level $x\geq 1/8$
the experimentally detected modulation
in the spin sector is characterized by $\epsilon_s \simeq 1/8$,
i.e. the magnetic ordering wavevector is ${\bf Q}_s^\ast \simeq \pi (3/4,1)$.
In this section, we shall investigate in detail the Fermi-surface reconstruction and the
arising Nernst signal as functions of various modulation strengths,
keeping ${\bf Q}_s$ fixed at ${\bf Q}_s^\ast$.
Wavevectors corresponding to longer modulation periods and doping $x<1/8$
will be discussed in Sec.~\ref{underdoped}.

By using the stripe-induced scattering potentials defined above, the quasiparticle dispersions
needed for a semiclassical calculation can be obtained by numerical diagonalization of the
Hamiltonian matrix.
The quasiparticle bands are spin degenerate because the paramagnetic (antiferromagnetic)
stripe-states are invariant under global spin-flips  (global spin-flips plus a
translation by one lattice spacing along the stripe-direction). Thus the spatially
averaged quantities, including the quasiparticle dispersions, cannot depend on the electron spin.
The general form of the Hamiltonian matrix for period-8 stripe order is (with ${\bf Q}_c^\ast=\pi(1/2,0)$)
\begin{widetext}
\begin{equation}
\left[
\begin{array}{cccccccc}
 \varepsilon_\bk & V_c^\ast & 0 & V_c & 0 & V_s^\ast & V_s & 0\\
 V_c & \varepsilon_{\bk+(\frac{\pi}{2},0)} & V_c^\ast & 0 & 0 & 0 & V_s^\ast & V_s\\
 0 & V_c & \varepsilon_{\bk+(\pi,0)} & V_c^\ast & V_s & 0 & 0 & V_s^\ast\\
 V_c^\ast & 0 & V_c & \varepsilon_{\bk+(\frac{3\pi}{2},0)} & V_s^\ast & V_s & 0 & 0\\
 0 & 0 & V_s^\ast & V_s & \varepsilon_{\bk+(\frac{\pi}{4},\pi)} & V_c^\ast & 0 & V_c\\
 V_s & 0 & 0 & V_s^\ast & V_c & \varepsilon_{\bk+(\frac{3\pi}{4},\pi)} & V_c^\ast & 0\\
 V_s^\ast & V_s & 0 & 0 & 0 & V_c & \varepsilon_{\bk+(\frac{5\pi}{4},\pi)} & V_c^\ast\\
 0 & V_s^\ast & V_s & 0 & V_c^\ast & 0 & V_c & \varepsilon_{\bk+(\frac{7\pi}{4},\pi)}\\
\end{array}
\right] \ .
\label{stripeorder}
\end{equation}
\end{widetext}
For brevity, in this matrix we dropped the momentum dependence in the scattering
potentials. Of course, these potentials in some cases depend on momentum, and this
dependence is easily obtained by labeling a potential connecting energies with momenta
${\bf k}+ {\bf q}$ and ${\bf k}+{\bf q}+{\bf Q}_{c/s}^\ast$ with the
momentum ${\bf k}+{\bf q}$ in the matrix~\eqref{stripeorder}.
In Fig.~\ref{antiphase}, the spin and charge distributions corresponding to both bond-centered and
site-centered period-8 stripe order are sketched.
Without loss of generality, we shall choose spin potentials with $V_s>0$.
Using the conventions given below Eqs.~(\ref{vcdw},~\ref{vsdw}) and $V_s$ being real,
it follows from the modulation of the chemical potential corresponding
to Fig.~\ref{antiphase} that $V_c(\bk) \equiv -V_c<0$ for site modulations,\cite{millisfoot} i.e.,
the $s$-wave part of the charge order.
Its $d$-wave part,\cite{vojta2008} described by bond modulations $\delta t$, will be chosen such that
sites with large spin density are connected by horizontal bonds (dimers), Fig.~\ref{antiphase}c,
which implies $\delta t > 0$.

\subsection{Fermi-surface reconstruction}

The particular geometry of the Fermi surface resulting from the diagonalization of
Eq.~\eqref{stripeorder} strongly influences the Nernst signal.
Typically, open electron orbits tend to give small contributions to the Nernst signal,
since they constrain the electronic motion mostly along one spatial direction
and lead to a small transverse flow of carriers, as we also checked numerically.
This can be understood from the expressions for the electrical and the thermoelectrical
Hall conductivity in Eq.~\eqref{boltzmann}. Their size is proportional to mass terms that
measure the band curvature, which tends to be small for open orbits as compared to closed orbits.

\begin{figure}
\includegraphics[width=3.7cm]{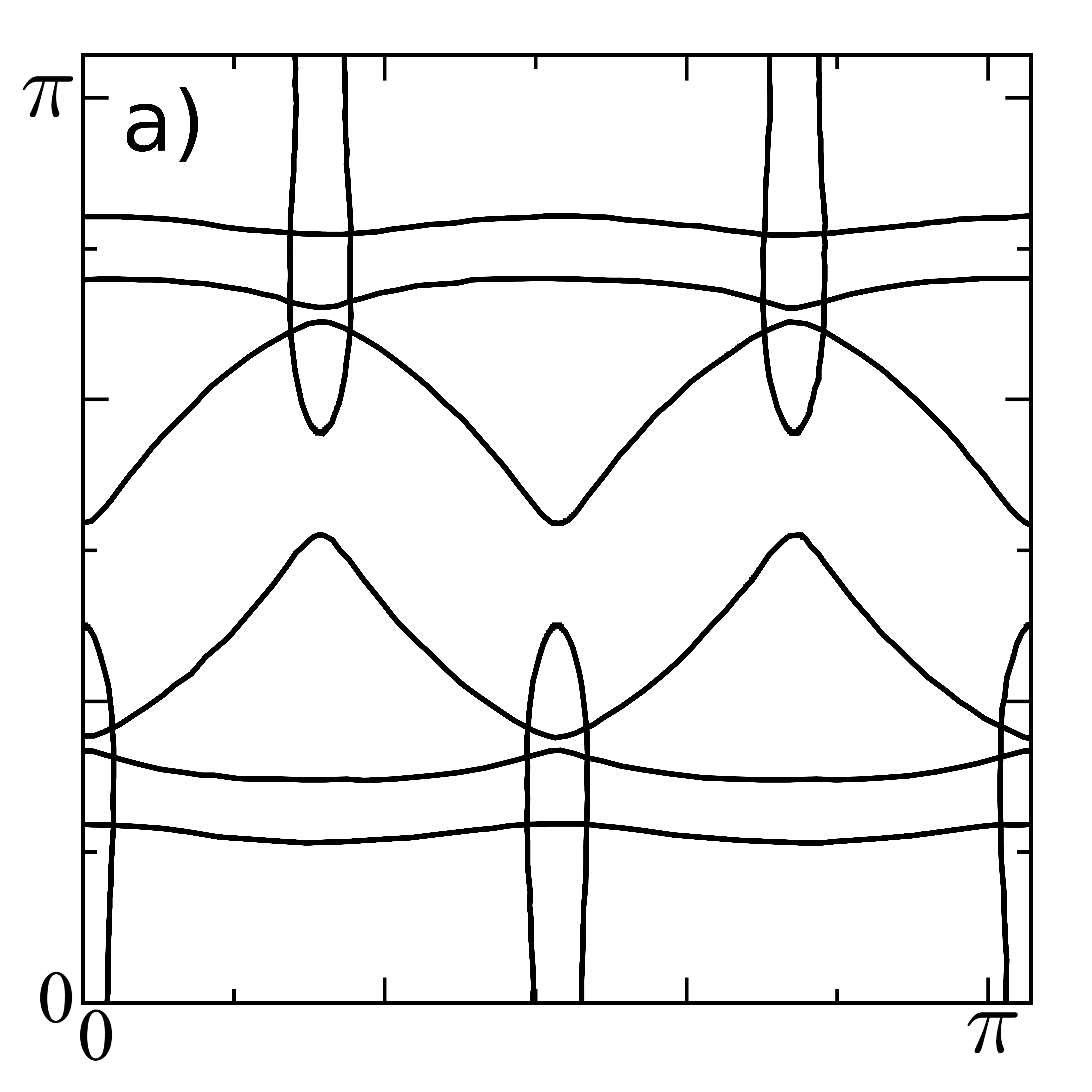} \includegraphics[width=3.7cm]{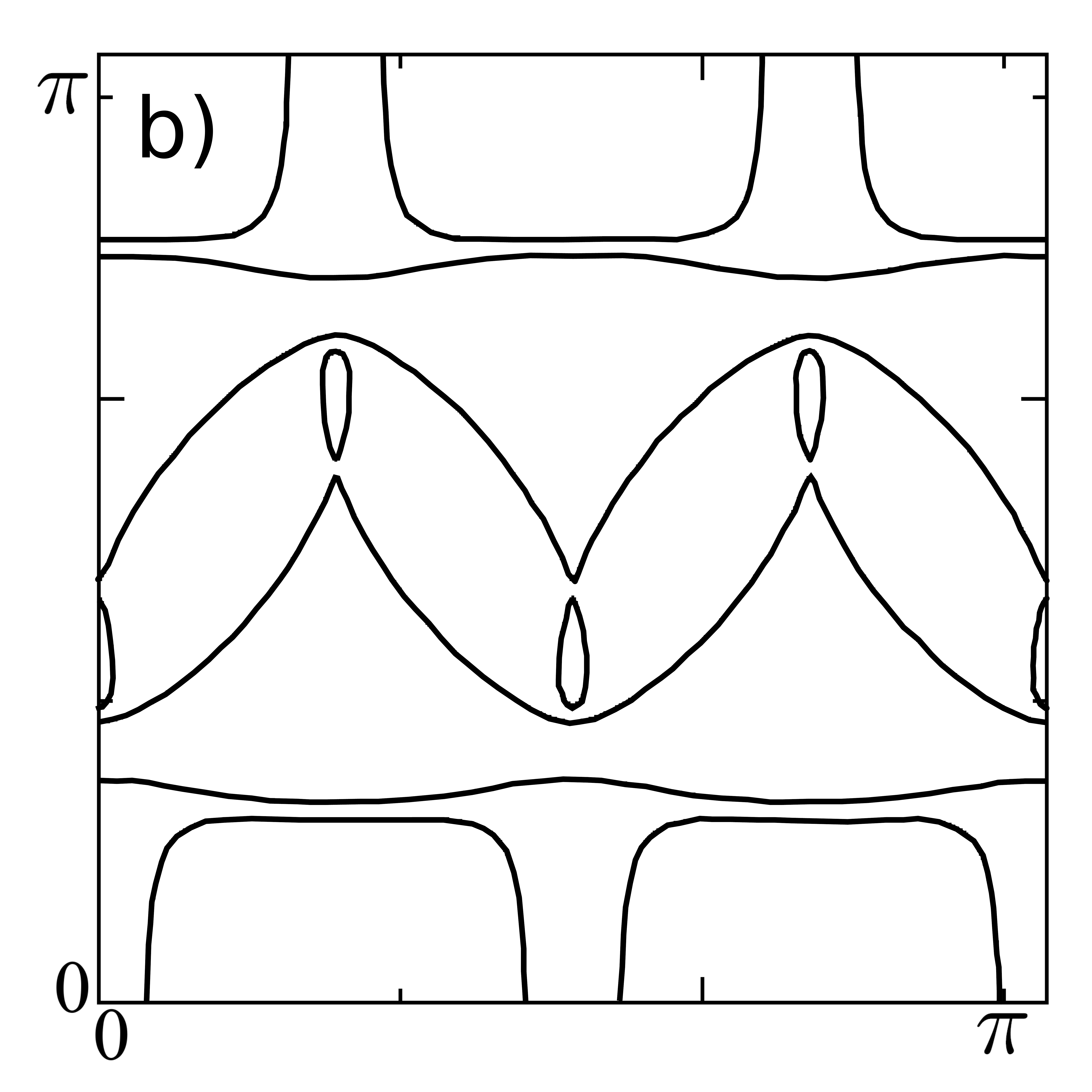}\\
\caption{
\label{purestripes}
Fermi surfaces for the bond-centered period-8 stripe states with
a) pure bond modulation, $\delta t=0.05$\,eV, and
b) pure spin modulation, $V_s=0.09$\,eV, plotted in the
first quadrant of the Brillouin zone of the underlying square lattice.
The Fermi surfaces are qualitatively equivalent to those obtained from site-centered spin or
charge potentials.
Without spin order (case a), besides open orbits only small
hole-like closed orbits with a large aspect ratio are present.
Spin order (case b) induces both hole-like and electron-like closed orbits.
}
\end{figure}

Concerning the Nernst signal as resulting from closed electron orbits,
a large Nernst signal resulting from quasiparticles usually requires the existence
of oppositely charged carriers, as it is strictly zero in the simple
Drude model as already noted by Sondheimer.\cite{sondheimer1948}
Generally, in any realistic system, such a cancellation will be incomplete.
As has been discussed in Ref.~\onlinecite{millis07,linmillis},
for the formation of closed electron orbits in the Fermi surface, a finite
spin-stripe potential is required, see Fig.~\ref{purestripes}.
Charge stripe order can only produce hole-like pockets
which eventually vanish in the limit of large charge stripe potential. Electron-like pockets
pinch of at the zone boundary in presence of finite spin stripe order, becoming smaller
upon increasing spin stripe potential.

\begin{figure}
\includegraphics[width=3.7cm]{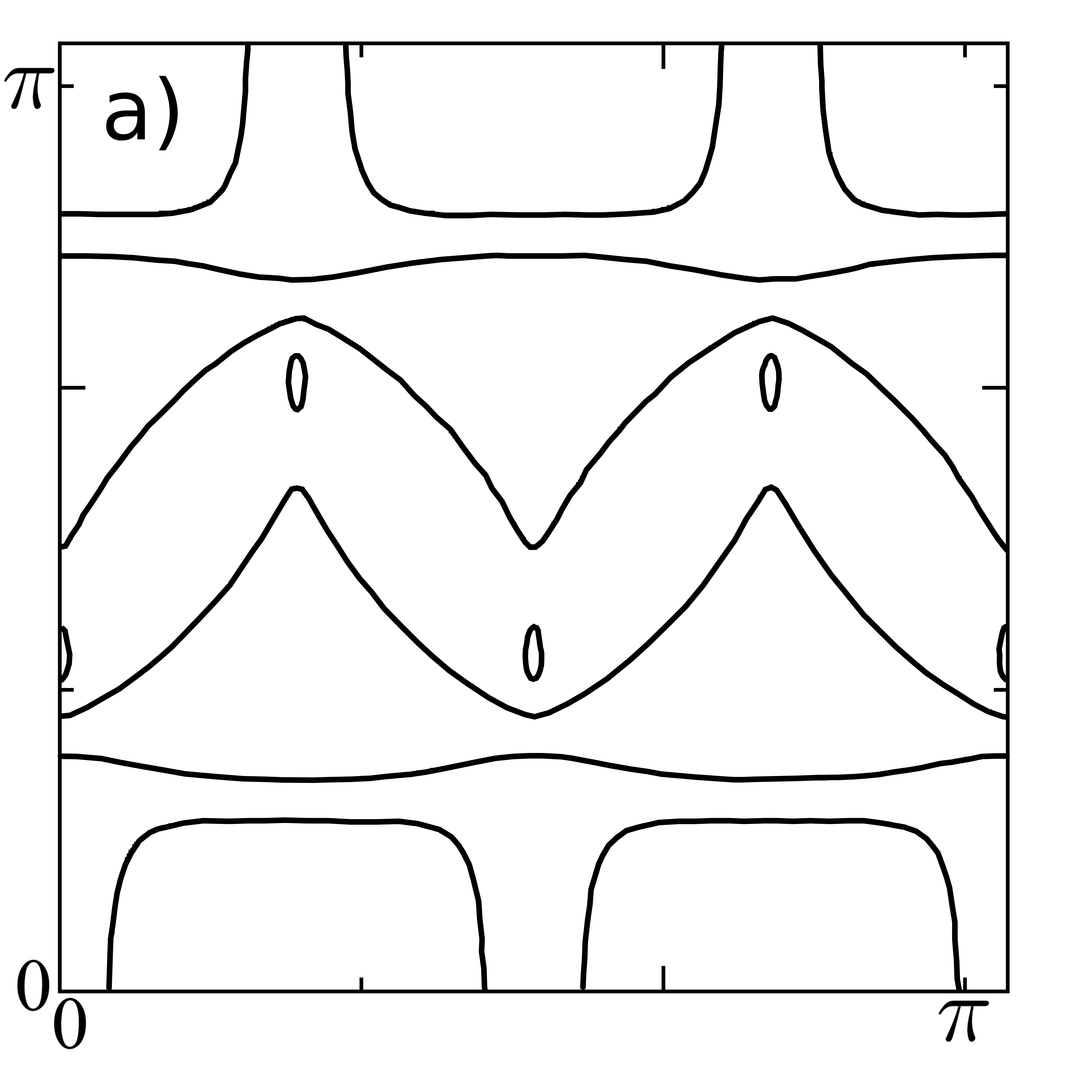}
\includegraphics[width=3.7cm]{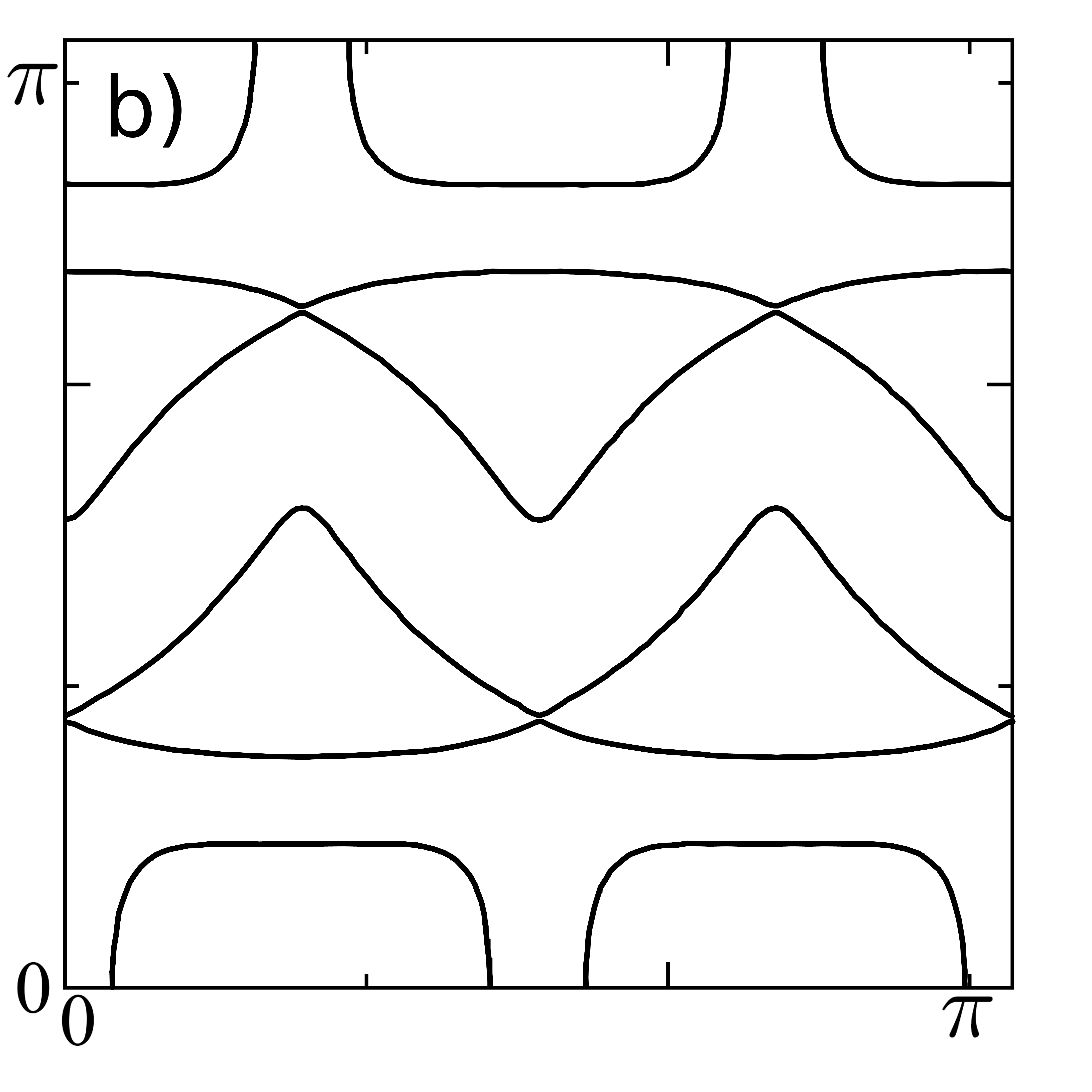}\\
\caption{\label{combinestripes}
Fermi surfaces for the bond-centered period-8 stripe states with combined spin and charge
modulation, plotted in the
first quadrant of the Brillouin zone of the underlying square lattice.
a) $V_s=0.09$\,eV, $\delta t=0.02$\,eV.
b) $V_s=0.09$\,eV, $\delta t=0.055$\,eV.
With increasing bond modulation, the small hole-like pockets shrink (case a)
and disappear (case b).
}
\end{figure}

\begin{figure}
\includegraphics[width=3.7cm]{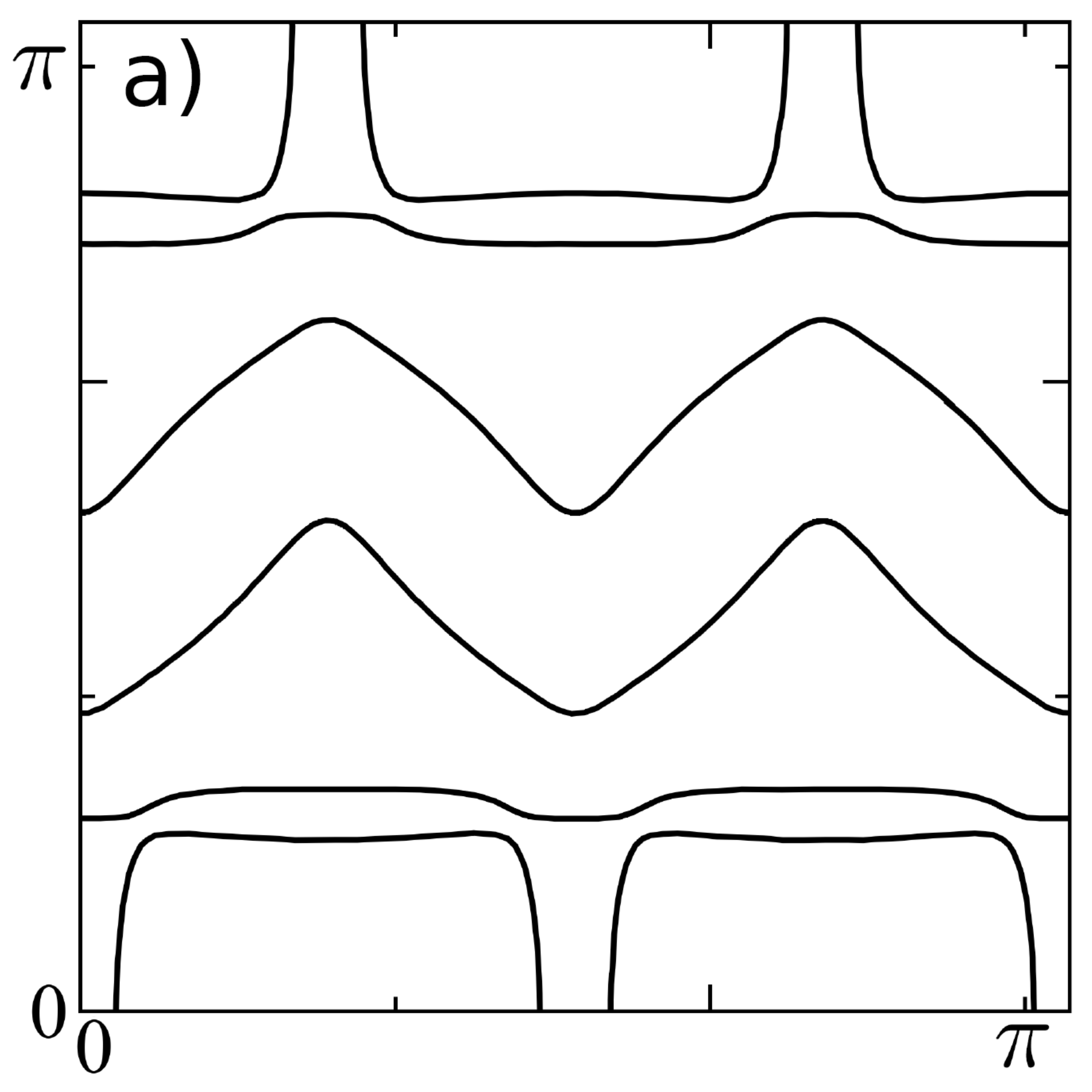} \includegraphics[width=3.7cm]{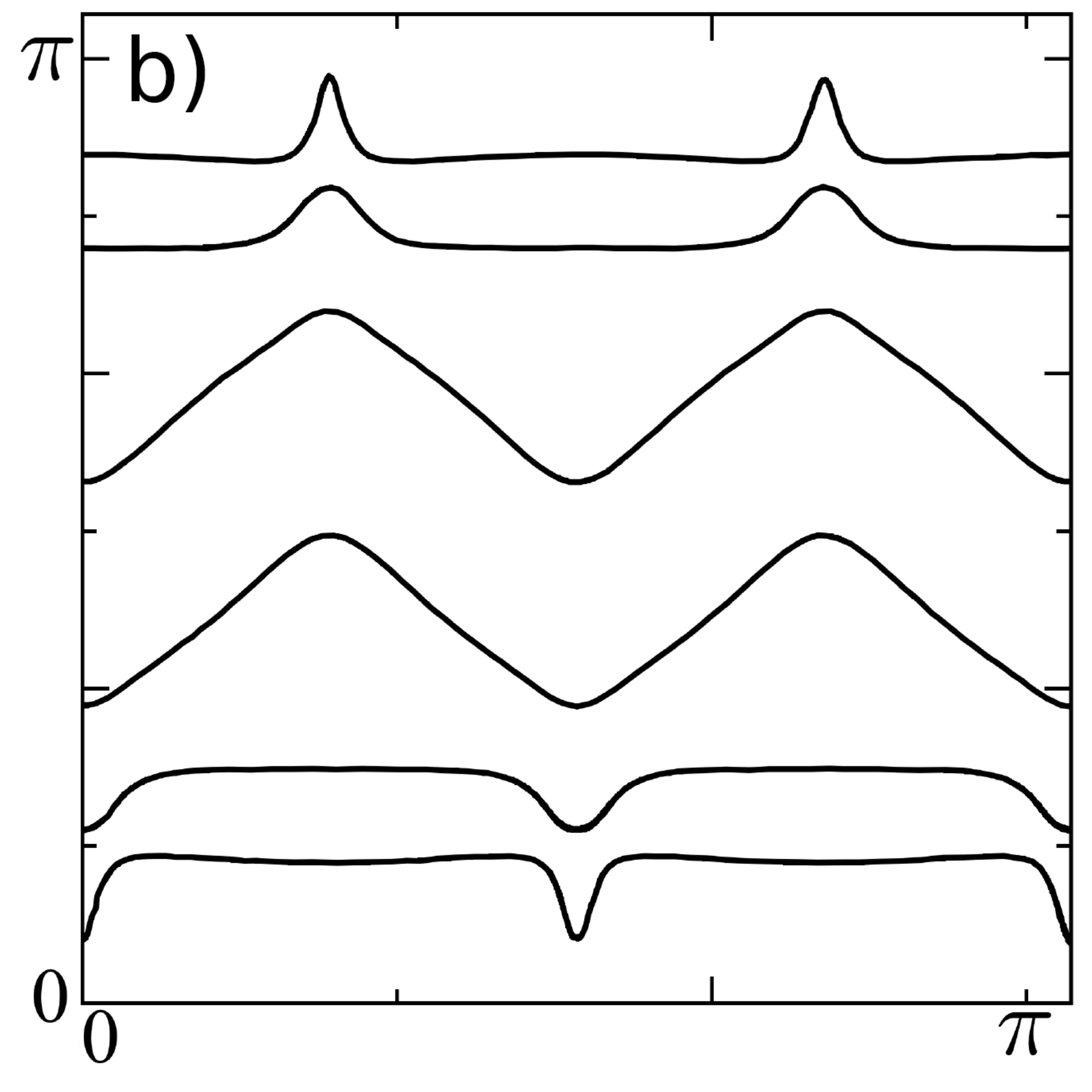}\\
\caption{
\label{sitesurface}
As in Fig.~\ref{combinestripes}, but for site-centered period eight stripe order.
a) $V_s=0.1$\,eV, $V_c=0.1$\,eV.
b) $V_s=0.1$\,eV, $V_c=0.15$\,eV.
As above, with increasing charge modulation the Fermi pockets disappear
in favor of open one-dimensional orbits.
}
\end{figure}

Thus, order in the spin sector seems crucial to produce a sizable Nernst signal.
These aspects motivate that we concentrate in the following on pure spin stripe order
(in the sense that charge order is only parasitic, see the discussion in Sec.~\ref{mf}).
Later on, we also study modifications
due to charge stripe order. The impact of charge order on Fermi surfaces as resulting from pure spin stripes
is illustrated in Fig.~\ref{combinestripes}. For very large charge potential, the electronic motion
is directed along the stripe direction, and closed electron orbits break
up even in presence of sizeable spin stripe potentials, as can be seen from Fig.~\ref{sitesurface}.

\subsection{Nernst effect from spin modulations}

Based on the above Fermi-pocket analysis,
we consider a situation of spin-driven stripe order first, i.e., our mean-field Hamiltonian
has modulated spin-dependent chemical potential as in Eq.~\eqref{vsdw}.
(For a modulation period of 8, this will induce weak charge order with period 4.)

\subsubsection{Nernst signal as function of modulation strength}

\begin{figure}
\includegraphics[width=7.0cm]{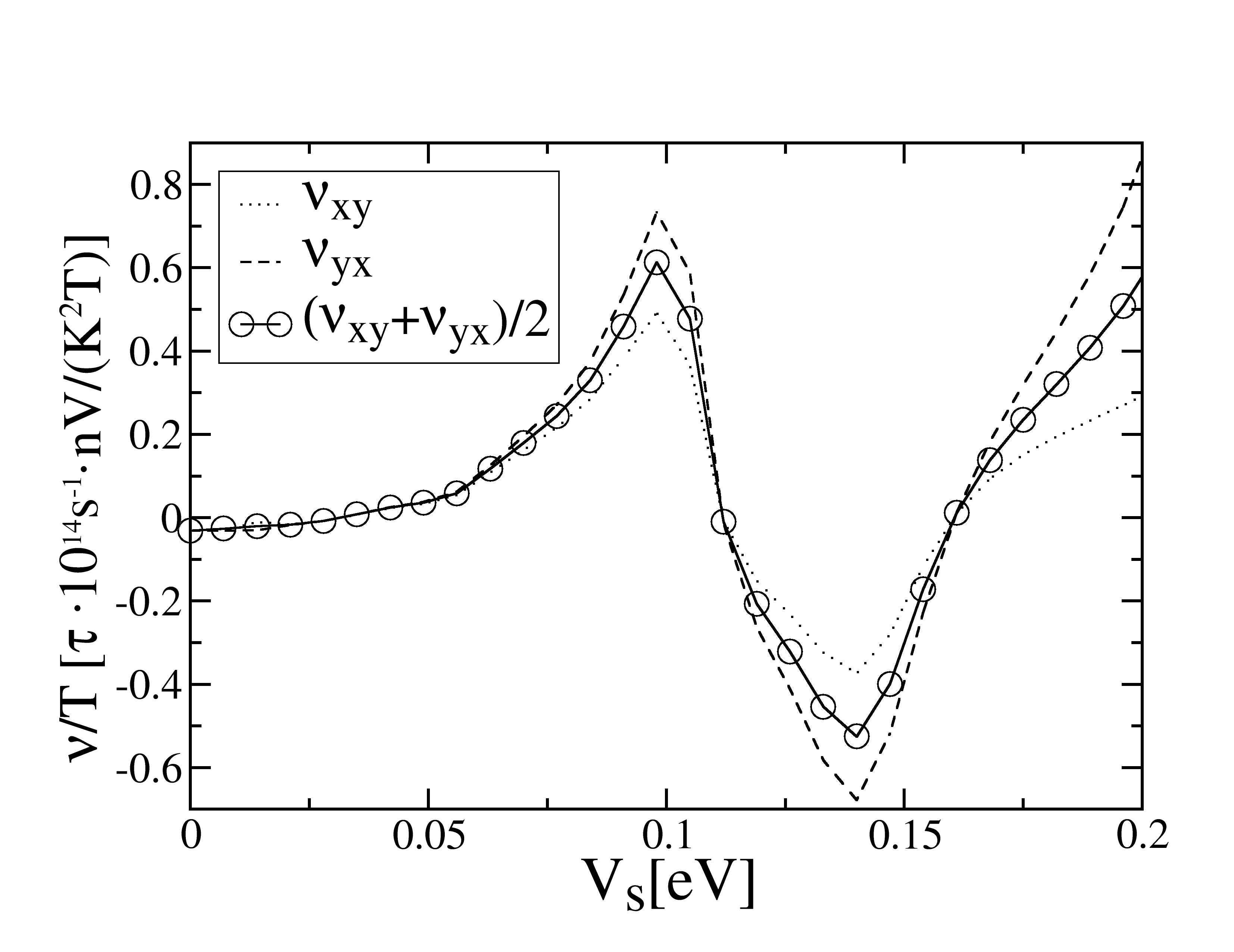}\\
\caption{\label{spinnernst}
Nernst effect for period-8 antiferromagnetic stripes at doping $x=1/8$ as function of
the spin modulation; the results are identical for the site-centered and bond-centered
cases.
The Nernst coefficient becomes negative at $V_s \simeq 0.1$\,eV, corresponding to maximal
local moments of $2\mu_B \langle S_z \rangle \simeq 0.3 \mu_B$.
Here and in the following, $\nu_{yx}$ is Nernst signal for $\vec{\nabla} T \parallel \hat x$.
The stripes have a modulation wavevector $\parallel \hat x$, i.e., run along $\hat y$,
such that $\nu_{xy}$ ($\nu_{yx}$) is defined with $\vec{\nabla} T$ parallel
(perpendicular) to the stripes.
}
\end{figure}

To set the stage,
we concentrate on the Nernst signal near $1/8$ doping, where the strong positive
enhancement is observed in experiments on \ndlsco.\cite{taill09}
Our result for the Nernst coefficient is shown in Fig.~\ref{spinnernst};
note that for our Hamiltonian the results for $\nu/T$ do not depend on whether the
spin stripes are site-centered or bond-centered, as the eigenvalues of the matrix
\eqref{stripeorder} do not depend on the complex phase of $V_s$ if $V_c=0$.
For small values of the spin potential, the Nernst coefficient is positive and highly
enhanced in comparison to the non-ordered state. For larger
spin-stripe potentials, the Nernst coefficient becomes negative and then again positive for
even larger spin stripe potentials. These changes can be traced back to the
stripe-induced changes of Fermi pockets: Upon increasing $V_s$, the small hole
pockets (see e.g. Fig.~\ref{purestripes}b) disappear at the maximum of $\nu/T$ in Fig.~\ref{spinnernst},
whereas the remaining open orbits split and form pockets at the minimum of $\nu/T$ (not shown).
The spatial anisotropy of the Nernst signal is small for all $V_s$.

To connect the parameter $V_s$ to experiments, the ordered magnetic moment may be used.
Experimentally, the maximum moment in the stripe structure at doping 1/8 in 214 compounds
has been estimated to be half of that of the undoped parent compound
(roughly $0.3 \mu_B$ or $\langle S_z \rangle = 0.15$),\cite{nachumi1998,stripe_rev2}
with different experimental techniques giving somewhat different results.
(It can be expected that the moment is smaller away from $x=1/8$.)
In \ybco, ordered magnetism in zero field is only observed for $y\leq0.45$,
but the order appears significantly field-enhanced.\cite{haug_mf}
(Based on the neutron-scattering and $\mu$SR data of Ref.~\onlinecite{haug_mf}
one may estimate the moment to be $0.05 \mu_B$ at zero field
and $0.07 \mu_B$ at 15\,T.)
In our mean-field calculation, we find that $\langle S_z \rangle_{max} = 0.15$ corresponds to
a scattering potential $V_s\simeq 0.1$\,eV for both bond and
site-centered stripes.
This value of $V_s$ is close to the maximum in the Nernst coefficient, and values of $V_s$ beyond this
maximum correspond to unrealistically strong magnetic order.

\subsubsection{Nernst signal as function of doping}

\begin{figure}
\includegraphics[width=7.0cm]{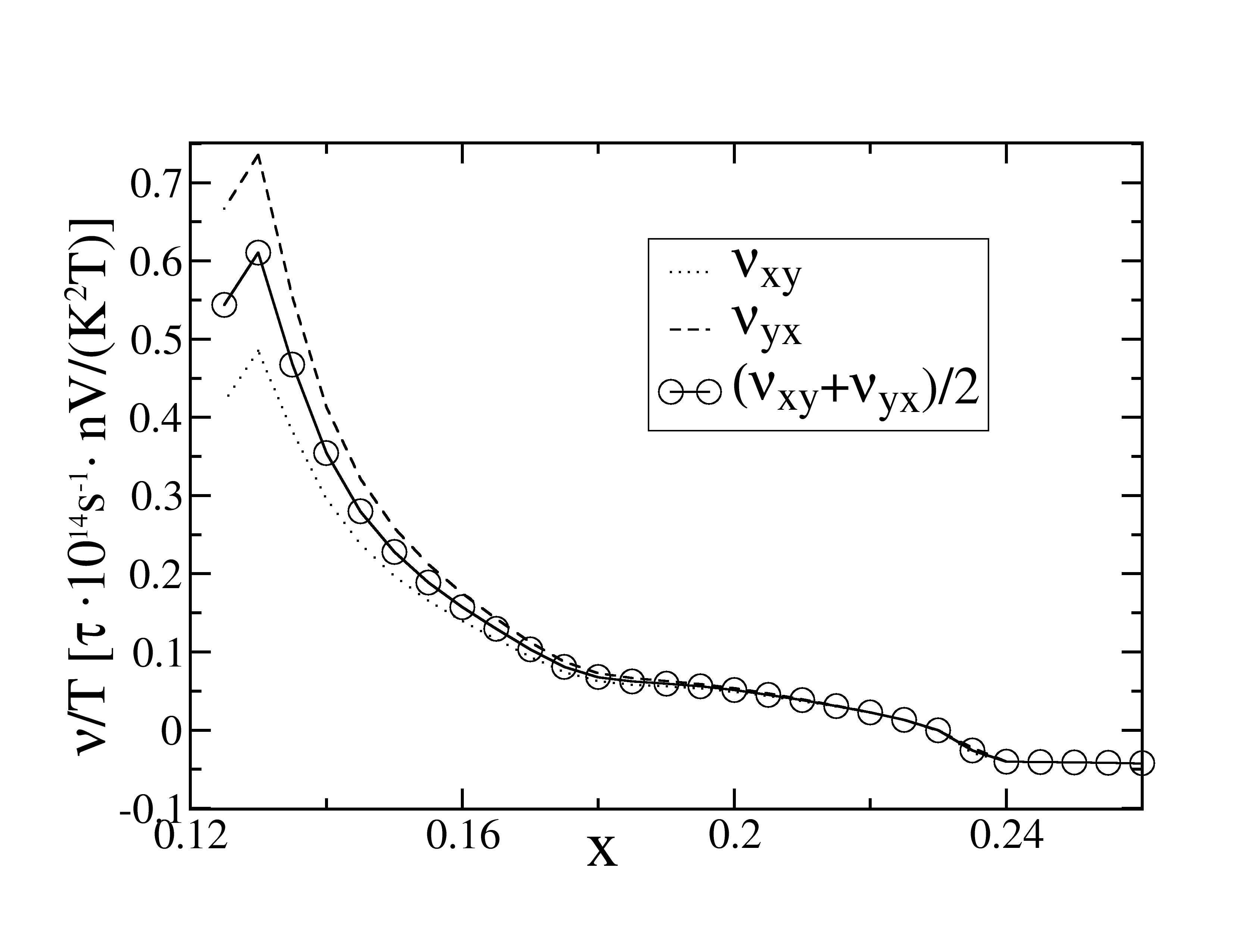}\\
\caption{\label{nernstdoping}
Doping dependence of the
Nernst coefficient for period-8 antiferromagnetic stripes, assuming a doping dependence
of the stripe order described by Eq.~\eqref{vsdop} and $V_0=0.15$.
It can be seen that the Nernst coefficient is similarly enhanced near $x=1/8$ for both types of
stripe order.
}
\end{figure}

We continue to study the doping dependence of the Nernst coefficient, for dopings $x\geq 1/8$
where the stripe period is doping-independent. Stripe order is maximally stable near $x=1/8$.
Experimentally, an extrapolation of the magnetic ordering temperature
in \ndlsco\ yields a critical doping $x_c=0.24$ where spin stripe order is suggested to
vanish.\cite{taill09,taillefer2009}
The simplest model assumption is then a mean-field dependence of the spin stripe order
parameter, $\phi_s \propto \sqrt{x-x_c}$ for $x<x_c$ at low $T$.
As the order parameter is linearly proportional to the modulation potential $V_s$,
we shall employ
\begin{equation}
V_s(x)=V_0\sqrt{1-x/x_c} \ ,
\label{vsdop}
\end{equation}
for $x$ below $x_c=0.24$ and $V_s=0$ elsewhere, while keeping the ordering wavevector
fixed at $\vec Q_s^\ast$.
The amplitude $V_0$ is set by the maximal local moment at $x =1/8$,
and we choose it such that  $\langle S_z \rangle_{max} = 0.15$ at this doping.
In Fig.~\ref{nernstdoping} we display the doping evolution of the Nernst coefficient
resulting from these assumptions, i.e., the doping axis in this figure corresponds to a
variation of both the band filling and the stripe amplitude.
As expected from the data in Fig.~\ref{spinnernst}, an enhanced positive Nernst
coefficient occurs over a large doping range, with a maximum at $1/8$ doping,
and little difference between site-centered and bond-centered spin stripes.
In the overdoped region, the Nernst coefficient becomes negative,
as is also observed in experiment.\cite{nernst_fl}
At lowest temperatures, the overall behavior agrees therefore well with the
experimental observations in \ndlsco.\cite{taill09}

Close to the critical doping $x \simeq x_c$, modifications of these results
due to magnetic breakdown have to be considered. If modifications of the band
structure by magnetic fields are neglected, the transmission amplitude is
analogous to Zener breakdown\cite{blount1962} and is given by the expression
\begin{equation}
\alpha = \exp\left[-\frac{\pi}{2}\frac{\Delta^2}{e\hbar B|v_x v_y|} \right] \ ,
\label{blountformula}
\end{equation}
where the Fermi velocities $v_x, v_y \simeq v_F$ are taken
at the related crossing point of the bare bands and $\Delta = 2V_s$
is the gap induced by SDW order.
Using the mean field dependence~\eqref{vsdop} of the SDW gap,
the doping range where the transmission amplitude is of $\mathcal{O}(1)$
is of the order
\begin{equation*}
\Delta x \simeq e \hbar v_F^2 B/(V_0^2)x_c \simeq 5.3  \times 10^{-3} B x_c \ ,
\end{equation*}
where we employed $V_0=0.15$\,eV and the universal Fermi velocity\cite{universal_vf}
$v_F=2.3\times10^7$ cm/s.
Considering magnetic fields of $\mathcal{O}(10\,\text{T})$, this doping range
is well separated from the important value $x=1/8$.

\subsubsection{Nernst signal as function of temperature}

We now turn to the temperature dependence of the Nernst coefficient.
In order to analyze how our quasiparticle calculation compares with experiment,
we model the effects of finite temperature by a temperature-dependent spin stripe potential
$V_0\sqrt{1-T/\Tsp}$, with $\Tsp \simeq 60$\,K at $x=1/8$ in \ndlsco\ as observed by neutron
scattering.\cite{ichikawa2000}
In addition, we model the temperature dependence of the quasiparticle scattering rate by
various parameterizations, {\em e.g.\/} by
the linear behavior $\tau^{-1}=a+bT$, with $b=a/70$\,K,
such that $\tau^{-1}(T\!=\!0) \simeq 2\tau_0^{-1}(T\!=\!70\,{\rm K})$.\citep{Valla2000}
(Here, $a\equiv\tau_0^{-1}$ remains a free parameter.)
Since the Nernst coefficient is proportional to the
relaxation time, this temperature dependence has no major influence on the overall shape
of the coefficient. Our numerical results show a peak in the Nernst coefficient at around $T=20$\,K.
Comparing this peak with the peak structure of height $50$\,nV/(KT) observed in Nernst
measurements in \eulsco,\cite{taill09} our calculation
requires a reasonable relaxation time $\tau \simeq 0.5\hbar / (k_B T)$ to reproduce
this peak height if the scattering rate is assumed
to be proportional to temperature, as observed experimentally in most parts of the Brillouin
zone.\cite{Valla2000}

\begin{figure}
\includegraphics[width=8.0cm]{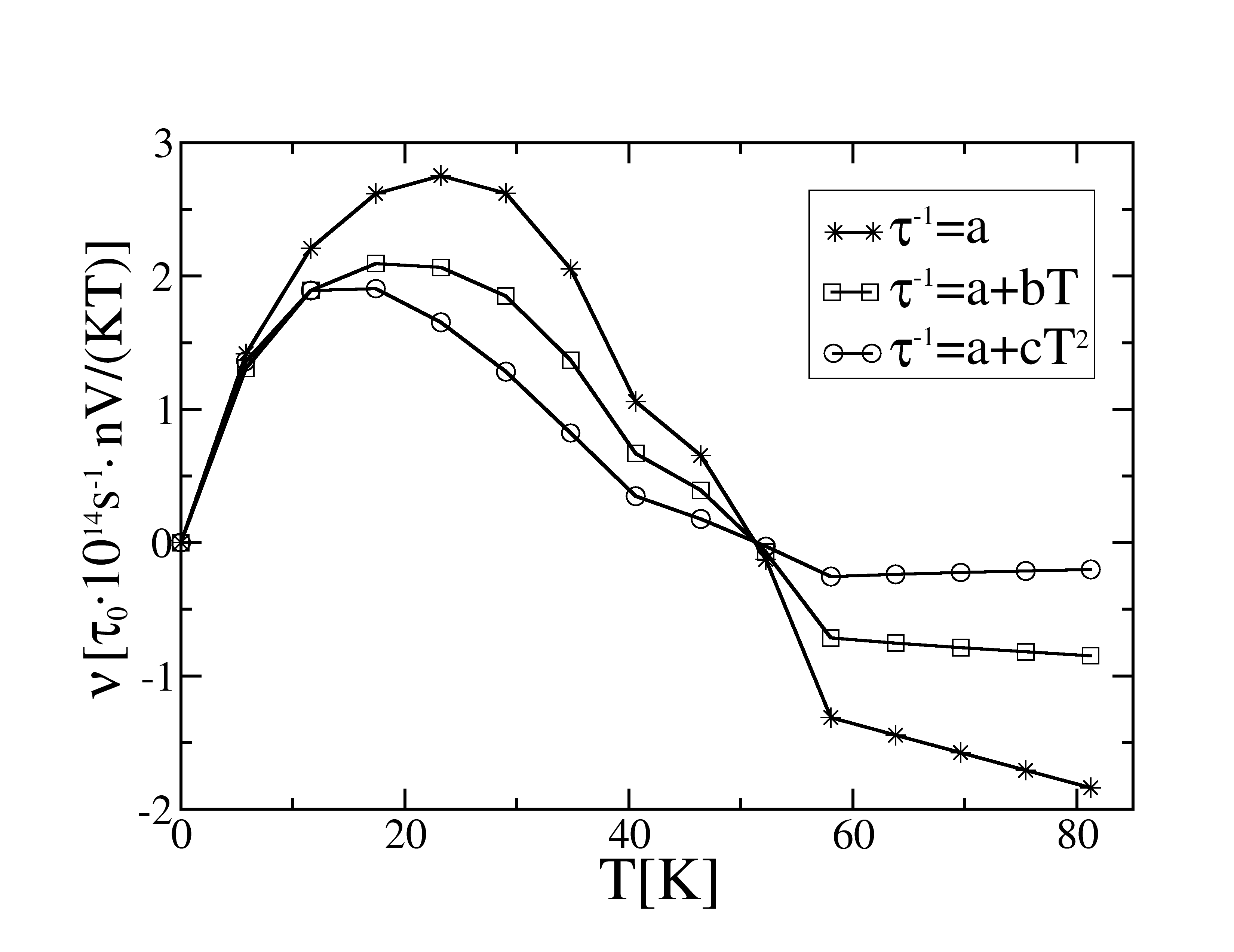}\\
\caption{
\label{temperaturenernst}
Temperature dependence of the Nernst coefficient for period-8 antiferromagnetic stripes. Upon increasing
temperature, the Nernst coefficient increases strongly to a large positive value which becomes
maximal at around 20\,K. Slightly below the ordering temperature $\Tsp\simeq60$\,K,
the coefficient becomes negative, as observed in experiment. The different scattering rates have been parameterized
with $a=\tau_0^{-1}$, $b=a/70$\,K and $c=a/800$\,K$^2$, and we set $V_0=0.1$\,eV}
\end{figure}

For a comparison to experiments, it has also to be considered that
a positive rise in the Nernst coefficient is already observed at twice the charge
ordering temperature, $T=2\Tch$.\cite{taill09}
Therefore, it appears that already stripe fluctuations can enhance the Nernst coefficient.

\subsection{Nernst effect from charge modulations}

Long-range static charge order has been observed mainly in 214 cuprates,
using neutron and x-ray scattering.\cite{stripe_rev1,stripe_rev2}
In addition, short-range static modulations in the charge sector have been detected on the surface of
underdoped \bsco\ and \cnco.\cite{kohsaka07}
However, reliable information about the amplitude of the charge modulation is lacking:
Most scattering experiments are not directly sensitive to the charge modulation, with
the exception of resonant soft x-ray scattering on \lbcoo\cite{abbamonte2005} whose
quantitative analysis (which gave a factor of 4 modulation of oxygen hole densities)
is, however, model dependent. From the STM data\cite{kohsaka07} one may infer a
typical modulation amplitude in the charge sector of $\pm 20 \hdots 30\%$,
if the contrast in the tunneling asymmetry is interpreted as density modulation.

Charge order (i.e. order in the spin-singlet sector) may exist without spin order, both
at $T=0$ and at finite temperatures.\cite{kivelson1998}
The latter is clearly seen e.g. in \eulsco\ in the temperature range between $\Tch\simeq 80$\,K and
$\Tsp \simeq 45$\,K near $x=1/8$.

\begin{figure}
\includegraphics[width=7.0cm]{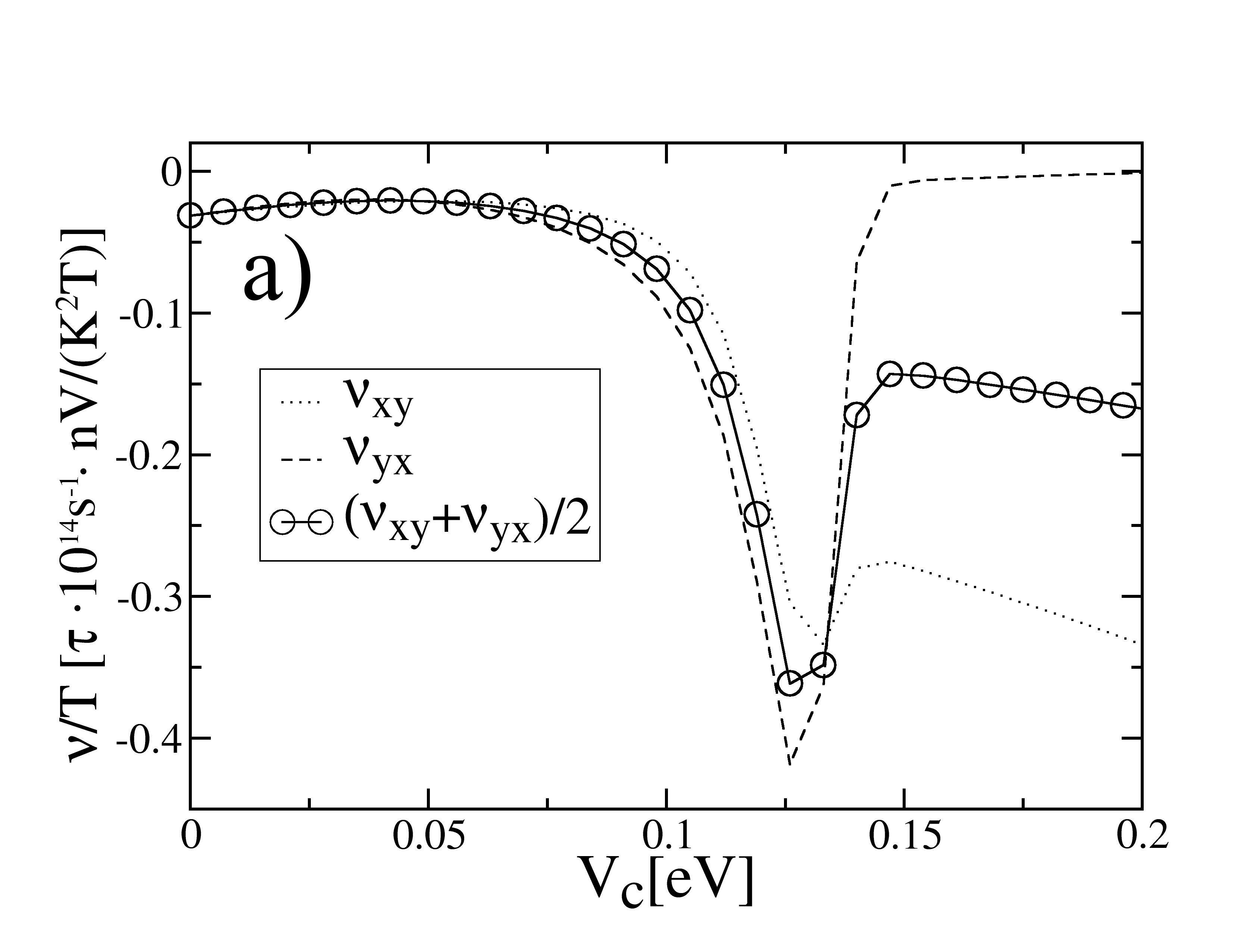}\\
\includegraphics[width=7.0cm]{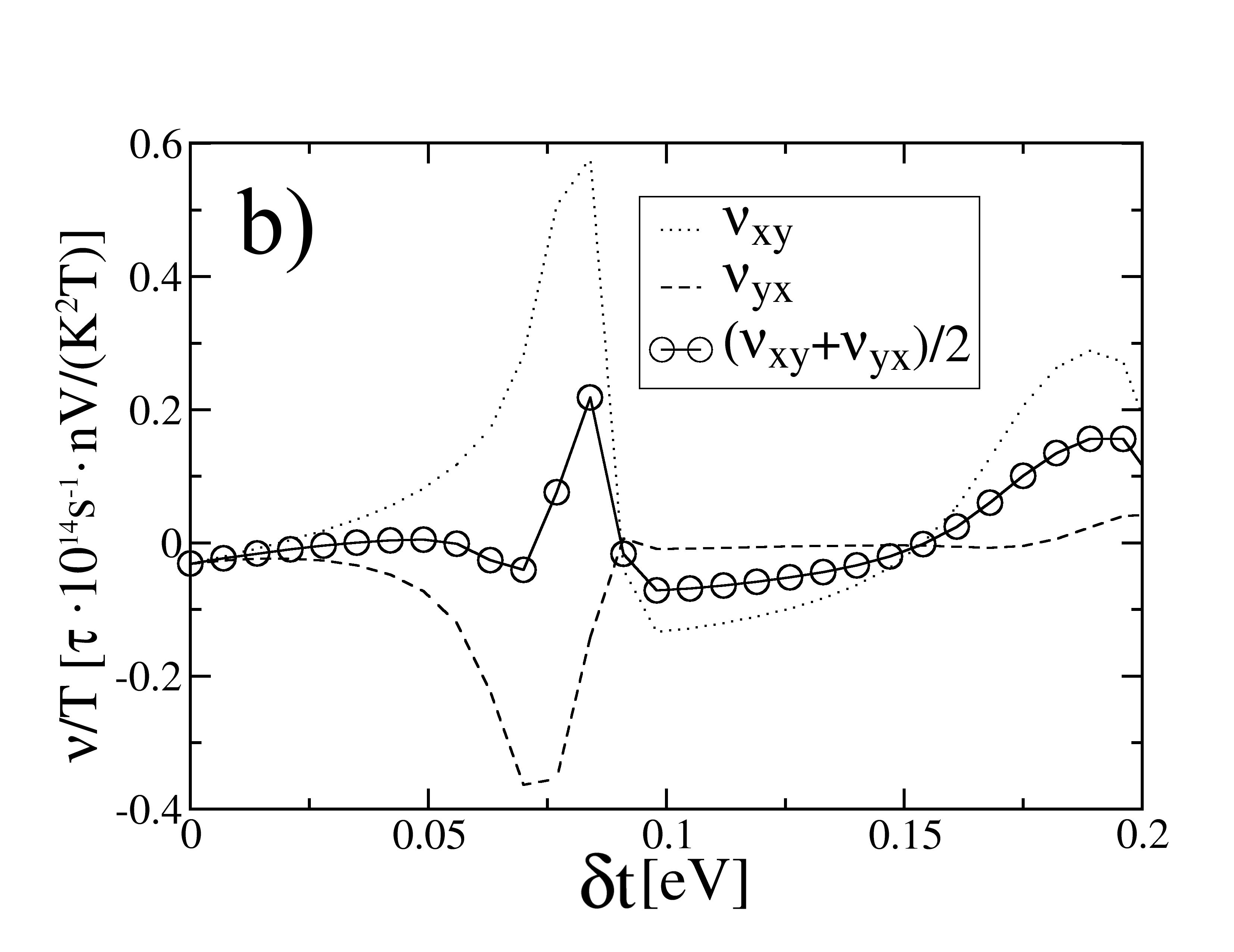}\\
\caption{\label{chargenernst}
Nernst effect for period-4 charge-only stripes at doping $x=1/8$ as function of
a) site-centered chemical-potential modulation and
b) bond-centered bond modulation.
The direction-averaged Nernst coefficient is clearly either negative or much less enhanced
than for spin stripe order for site-centered stripe order.
In addition, it is small everywhere where modulation in the charge channel does not exceed $30\%$,
corresponding to $\delta t\lesssim 0.06$\,eV and $V_c\lesssim 0.1$\,eV.
The large anisotropy in panel b is due to the presence of extremely elongated hole
pockets.
}
\end{figure}

In this subsection, we consider the effect of charge-only modulations on the Nernst coefficient. As
discussed in Sec.~\ref{mf}, order in the charge sector may be described by modulated
on-site potentials for site-centered stripes or by a spatially modulated hopping
amplitude (describing bond order) in the case of bond-centered stripes.
Sample results for the Nernst coefficient are shown in Fig.~\ref{chargenernst}.

A few remarks are in order. First, charge order with charge modulation below 30\%
cannot produce closed electron orbits, as shown in Fig.~\ref{purestripes},
and only hole-like orbits emerge.
For site-centered stripes, this was already stated in Ref.~\onlinecite{millis07}.
The direction-averaged Nernst coefficient shown in Fig.~\ref{chargenernst}
is negative (or positive, but small) for both site-centered and bond-centered charge order.
Overall, the magnitude of the signal is also rather small for reasonable potential strengths
($\delta t=0.055$\,eV leads to $20\%$ ($30 \%$) modulation  of vertical (horizontal) bond density,
while site-centered stripes with $V_c=0.1$\,eV lead to $30\%$ modulation of charge density).
Thus, it cannot account for the positively enhanced Nernst coefficient which has been measured
in presence of stripe order.\cite{taill09}
One interesting feature of Fig.~\ref{chargenernst}b is the large anisotropy of $\nu$ in
the range $\delta t \approx 0.06\ldots0.08$\,eV. This can be traced back to elongated
hole pockets as in Fig.~\ref{purestripes}a which exist in this parameter range.
Everywhere else the Nernst anisotropy is moderate or small.

\subsection{Combined spin and charge modulations}

\begin{figure}
\includegraphics[width=7.0cm]{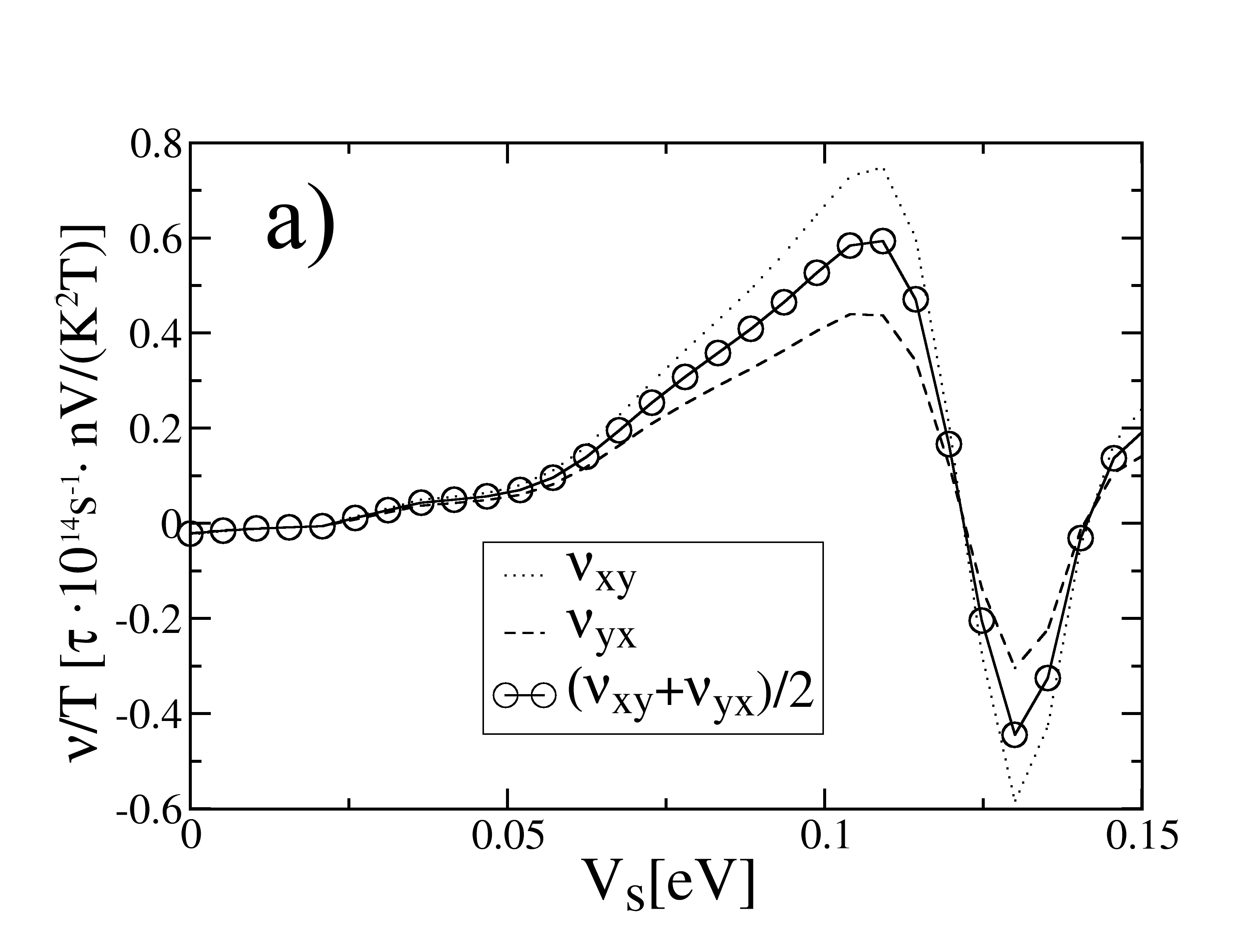}\\
\includegraphics[width=7.0cm]{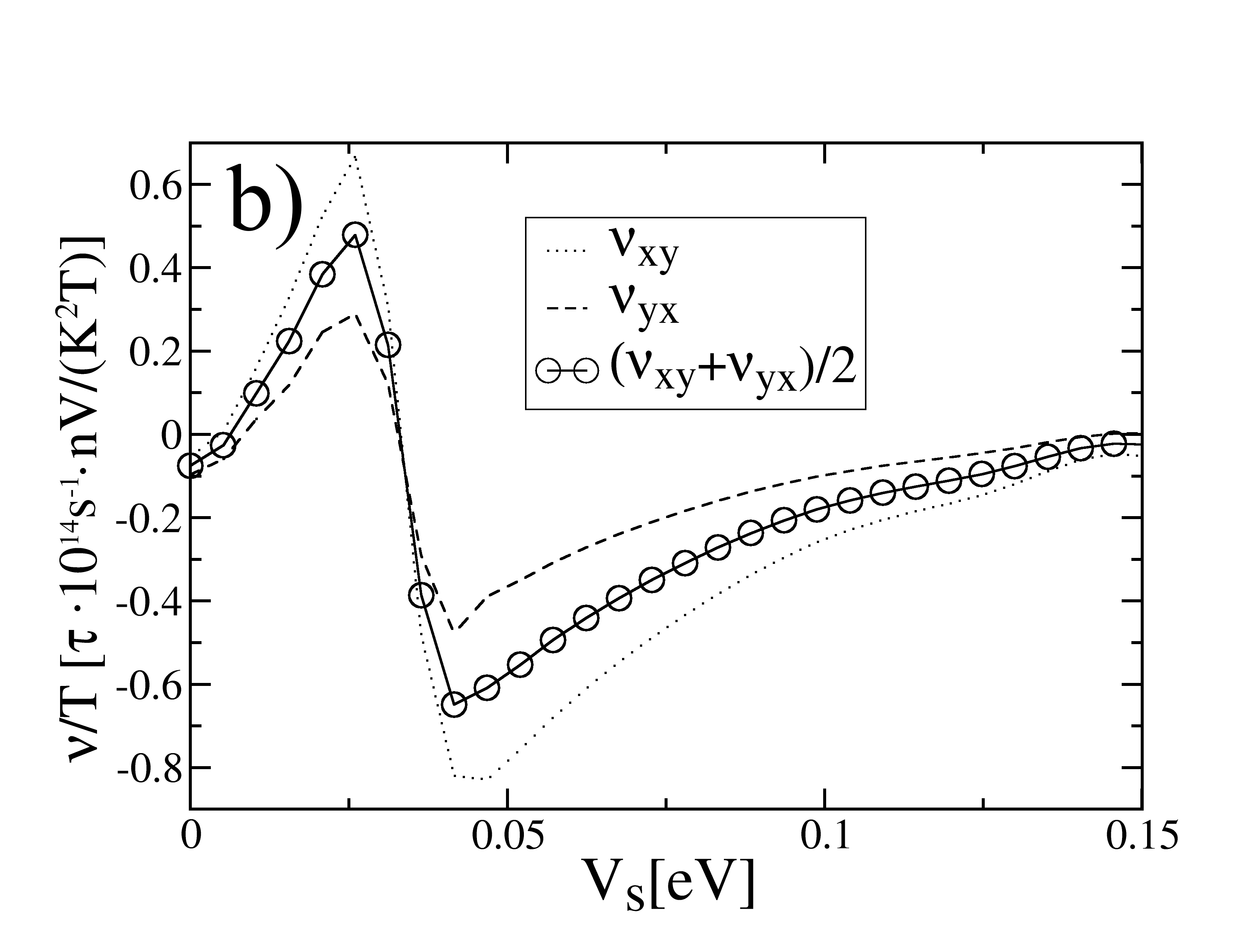}\\
\includegraphics[width=7.0cm]{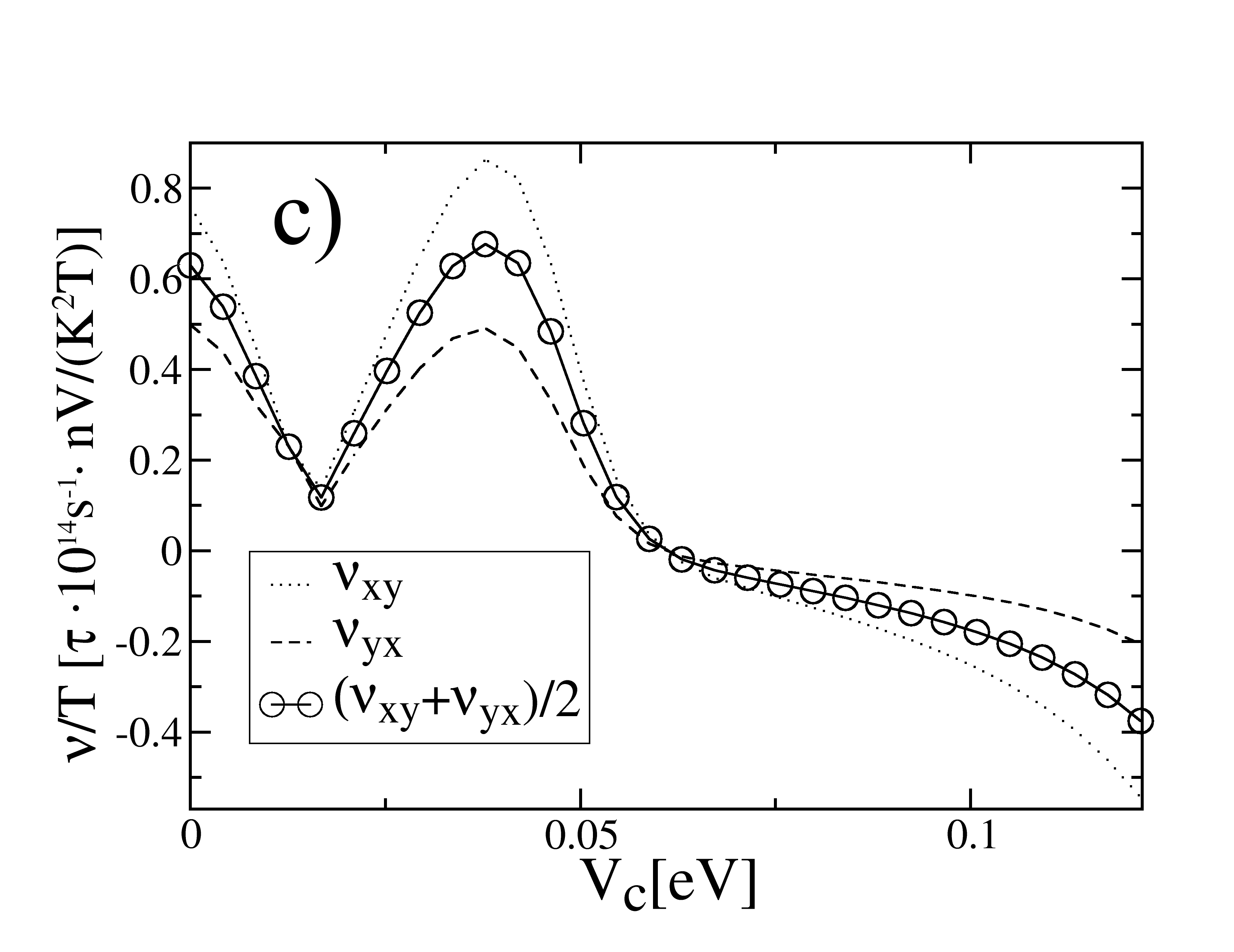}\\
\caption{\label{combinednernst}
Nernst effect for site-centered period-8 stripes with combined spin and charge order.
a) Fixed $V_c=0.03$\,eV as a function of $V_s$.
b) Fixed $V_c=0.1$\,eV as a function of $V_s$.
c) Fixed $V_s=0.1$\,eV as a function of $V_c$.
For a spin stripe potential of $V_s=0.1$\,eV, charge potentials above the moderate value $V_c=0.05$\,eV
lead to a negative or small Nernst coefficient, see panel c).
}
\end{figure}

\begin{figure}
\includegraphics[width=7.0cm]{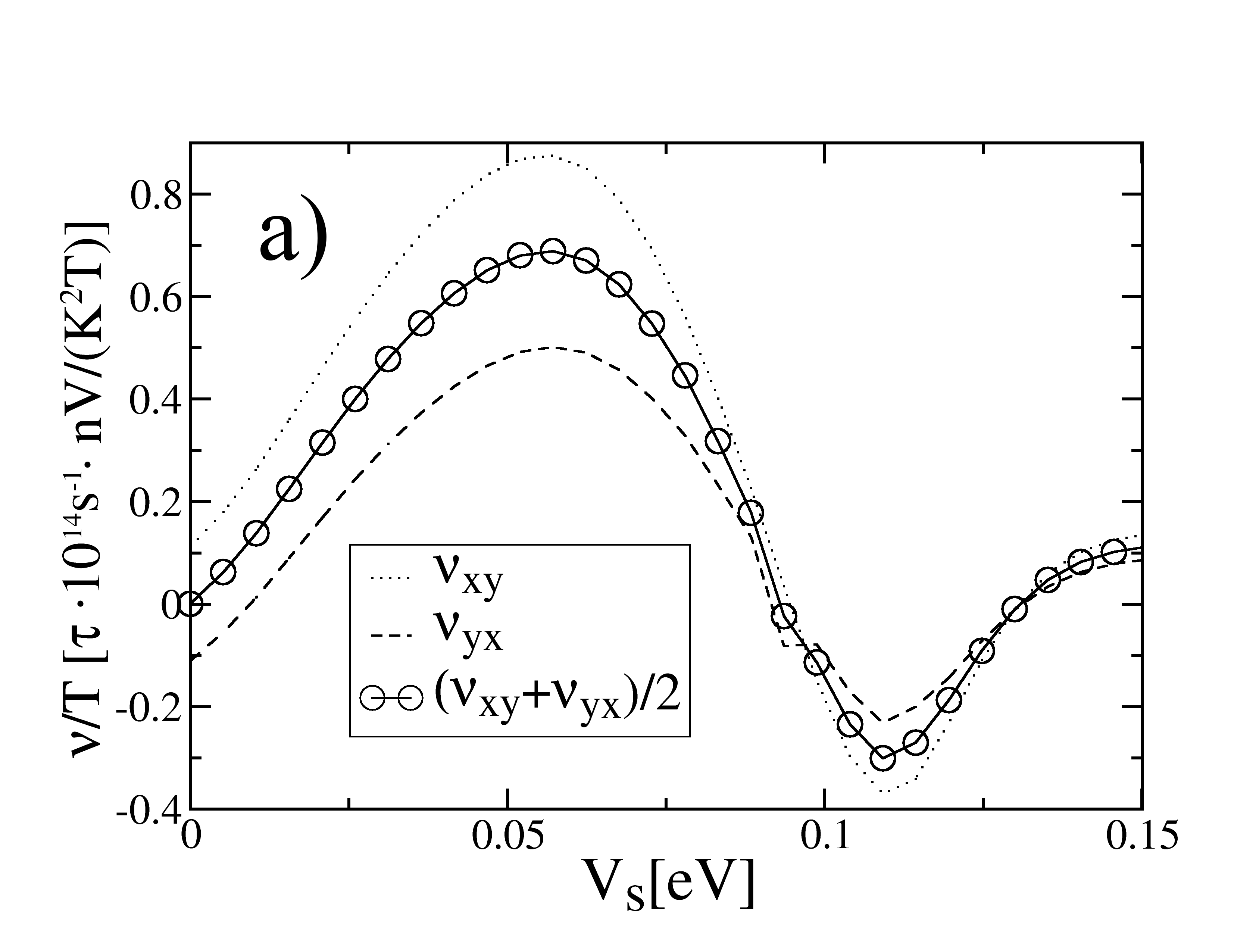}\\
\includegraphics[width=7.0cm]{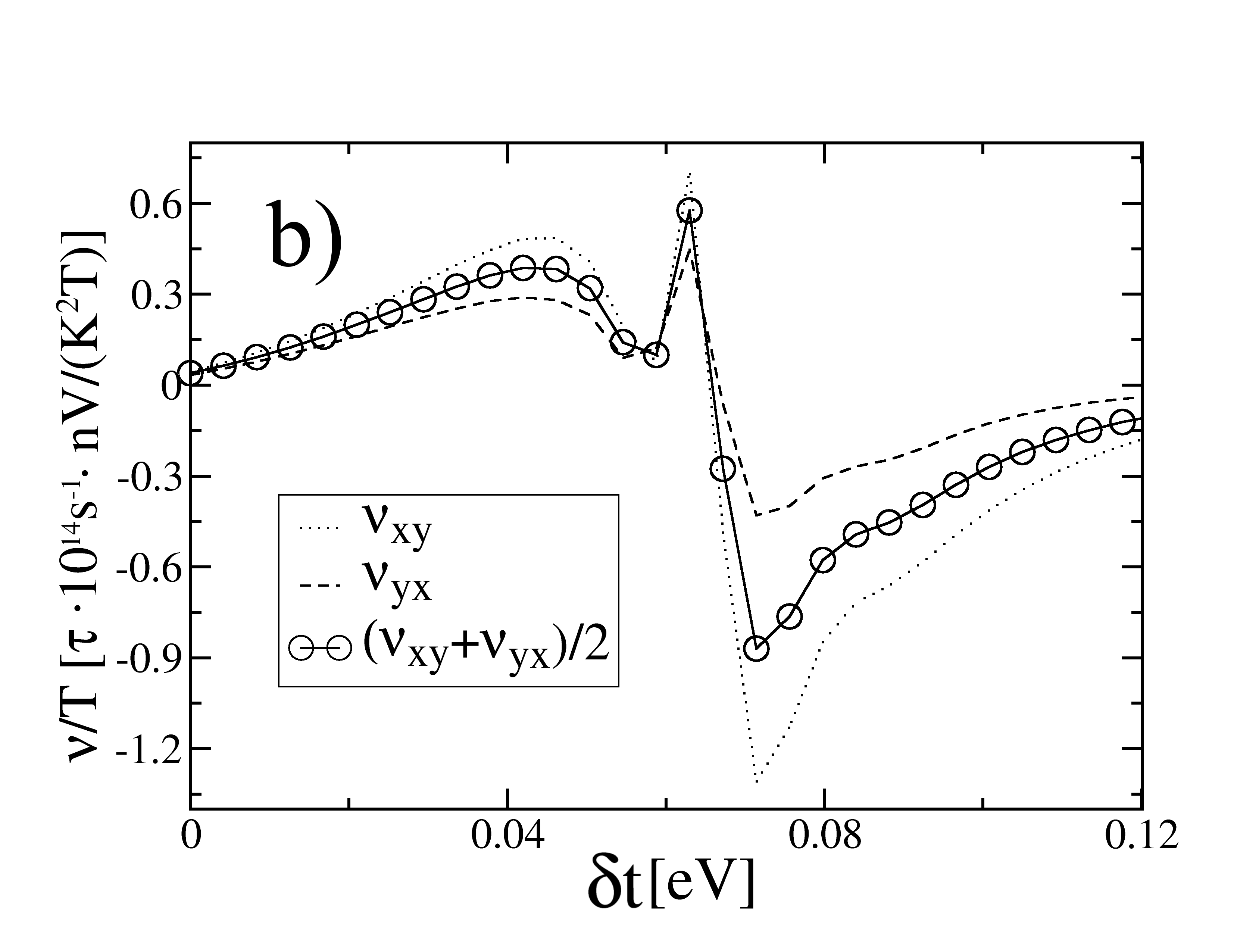}\\
\caption{\label{stripenernst}
As in Fig.~\ref{combinednernst}, but for bond-centered period-8 stripes.
a) Fixed $\delta t=0.055$\,eV as a function of $V_s$.
b) Fixed $V_s=0.09$\,eV as a function of $\delta t$.
As is depicted in panel a), for a wide range of spin stripe potentials
below $V_s \simeq 0.09$\,eV the Nernst coefficient is positively enhanced.
($V_s=0.1$\,eV  corresponds to an ordered moment of $\simeq 0.3\mu_B$).
For bond modulations $\delta t \lesssim 0.05$\,eV, the Nernst coefficient can
remain positive, see panel b).
}
\end{figure}

We are thus lead to consider the effects of combined spin and charge stripe order.
Adding charge order on top of spin stripe order has the effect of breaking up closed electron orbits into open orbits
for sufficiently strong charge order, see Fig.~\ref{sitesurface}.
It is therefore natural to expect that transport properties resulting from
pure spin stripe order will qualitatively change if charge stripe order becomes too strong.
For on-site modulations,
a quantitative measure for charge modulation is the relative local deviation
from the mean conduction electron density. In the site-centered case, a deviation of
$20 \%$ corresponds to $V_c=0.07$\,eV  in presence of a spin potential of $V_s=0.1$\,eV.
It turns out that the Nernst coefficient remains strongly enhanced for charge potentials
of up to about $V_c=0.05$\,eV , while the coefficient becomes very small or negative for
stronger charge potentials, see Fig.~\ref{combinednernst}c.
This behavior would therefore be compatible with the normal-state Nernst coefficient
in \ndlsco\ if charge order leads only to
modulations of $15\%$ or below in the charge sector.
A similar behavior is obtained for bond-centered spin stripes with additional bond modulations,
shown in Fig.~\ref{stripenernst}.
For a strong bond modulation of $\delta t=0.055$\,eV with a kinetic energy modulation
of about $20-30\%$, the Nernst coefficient is negative only in a small range of spin stripe potential, Fig.~\ref{stripenernst}a.
Finally, if the spin stripe potential is larger than $V_s=0.1$\,eV  (as is required
to produce a maximal local moment of $0.2 \mu_B$ or more), the Nernst coefficient is positive
also for the large bond modulation of $\delta t=0.055$\,eV .
In order to account for the observed positive normal-state Nernst coefficient\cite{taill09}, this behavior
suggests rather a bond-centered nature of charge order
in \ndlsco\ if the modulation in the charge sector exceeds $15 \%$.


\section{\label{underdoped} Nernst effect below doping $x=1/8$}

The underdoped regime of the cuprates with hole dopings below $x=1/8$
is of interest for various reasons. First of all, the ordering wave
vector in stripe-ordered 214 compounds is strongly doping dependent,
$\epsilon_s\simeq x$,
in contrast to the constant modulation period observed for $x\geq 1/8$.
In addition, recent high magnetic field experiments on underdoped \ybco\
at hole doping of about $x=0.1$ have reported quantum oscillations,
interpreted in terms of multiple small Fermi pockets\cite{doiron07,LeBoeuf07}
and a negative Hall coefficient.\cite{LeBoeuf07}

Very recently, also a negative normal-state Nernst coefficient has been reported in underdoped
\ybco,\cite{daou2009} which, moreover, was found to display a strong spatial
anisotropy.\cite{taill10b}
We note that in \ybco, tendencies toward stripe order appear weaker than in 214
materials. While incommensurate low-energy spin fluctuations have been observed over
a large doping range of \ybco, which become static around $y=6.45$,
there is no clear-cut evidence for charge order in these materials.

In the following, we consider two cases of stripe order
with collinear spin order of periods 10 and 16 in order to analyze the normal-state
Nernst coefficient corresponding to far underdoped samples.
Period 10 is motivated by the doping level $x=0.1$ where quantum oscillations have been reported,
period 16 is motivated by the neutron scattering work on \ybco\ with $y=6.45$
where incommensurate correlations at $\vec{Q}_s = 2\pi(0.5\pm\epsilon_s,0.5)$ with $\epsilon_s \simeq
0.06$ were detected.\cite{haug_mf,hinkov08}
As before, we will neglect the interlayer hopping part of the dispersion
as well as effects of bilayer splitting and the ortho-II potential.
(Note that various experiments have been performed on non-ortho-II ordered samples, e.g. quantum
oscillations have been reported for such samples.\cite{singleton2009})

\subsection{Period-16 stripe order}

Following Ref.~\onlinecite{harrison09}, we will approximate the
experimentally detected \cite{haug_mf,hinkov08} incommensurability
$\epsilon_s=0.06$ by the rational value $1/16$ in order to obtain the reconstructed Fermi
surface from the eigenvalues of a finite  matrix. In this approximation, gaps of order
$2\Delta_m \sim 2V_s^m/t^{m-1}$ with $m \gg 1$ are neglected. For experimentally relevant
field strengths of $10$\,T or more, these gaps are broken through
if $V_s \ll t \sim t_1$
and can indeed be neglected. This is especially the case for the ratio $V_s=t_1/6$ used
in Ref.~\onlinecite{harrison09}, for which the transmission amplitude
through the $m=3$ gap in $B=20$\,T is $\simeq 94.1 \%$ (according to formula~\eqref{blountformula},
using $v_F=2.3\times 10^7 \text{cm/s}$ \cite{universal_vf}).
In addition, we neglect also all other gaps with $m>1$.
These are either broken through by magnetic breakdown for $m>2$ or they do not lead to
closed orbits ($m=2$), as discussed in Ref.~\onlinecite{millis07} .

Including both spin and charge order to our modelling leads to the $16 \times 16$
Hamiltonian matrix
\begin{equation}
H=\left(
\begin{array}{cccccc}
 \varepsilon_{\bf k} & V_s^\ast & V_c^\ast & \hdots & V_c & V_s \\
 V_s & \varepsilon_{\bk+{\bf Q}_s} & V_s^\ast & \hdots & 0 & V_c \\
 V_c & V_s & \varepsilon_{{\bf k}+2{\bf Q}_s} & \hdots & 0 & 0 \\
 \vdots & \vdots & \vdots & \ddots & \vdots & \vdots \\
 V_c^\ast & 0 & 0 & \hdots & \varepsilon_{{\bf k}+14{\bf Q}_s} & V_s^\ast \\
 V_s^\ast & V_c^\ast & 0 & \hdots & V_s & \varepsilon_{{\bf k}+15{\bf Q}_s} \\
\end{array}
\right) \ .
\label{16bands}
\end{equation}
Again, momentum dependence of the scatttering potentials has been dropped in
Eq. (13.9) and can be restored by labeling a potential connecting energies with momenta
${\bf k}+ {\bf q}$ and ${\bf k}+{\bf q}+{\bf Q}_{c/s}^\ast$ with the
momentum ${\bf k}+{\bf q}$ in the matrix~\eqref{16bands}.
\begin{figure}
\includegraphics[width=7.0cm]{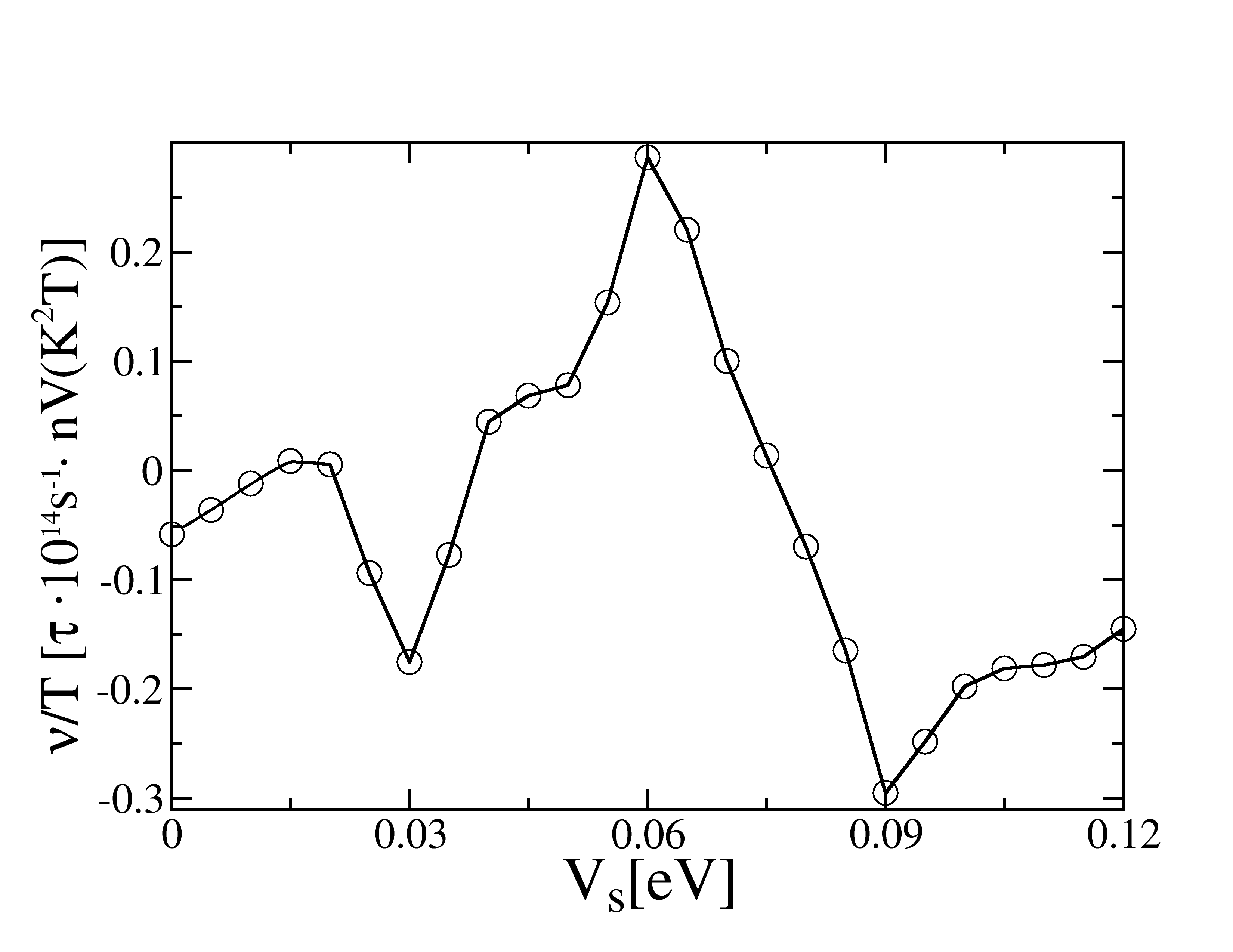}\\
\caption{\label{ybconernst}
Nernst coefficient $\nu_{yx}$ for a period-16 SDW order as a function
of $V_s$ with $x=0.1$.
For $V_s \gtrsim 0.07$\,eV (corresponding to a maximal local moment of $m \gtrsim 0.20 \mu_B$)
the Nernst coefficient turns negative with an enhanced
amplitude in comparison to the non-ordered state.
Note that in a stripe (or SDW) picture for \ybco, the stripes run along the b axis
(as inferred from neutron scattering\cite{haug_mf,hinkov08})
and our $\nu_{yx}$ corresponds to the Nernst signal with $\vec\nabla T$ along the a axis.
}
\end{figure}
Results for the Nernst coefficient of period-16 spin stripe order are shown in Fig.~\ref{ybconernst}.
The modulation parameter $V_s$ can again be connected to the magnitude of ordered moment.
As stated above, the experimentally detected moment increases from $0.05\,\mu_B$ at zero field
to $0.07\,\mu_B$ at 15\,T,\cite{haug_mf} which suggests that in field of 50\,T as applied in quantum
oscillation measurements an ordered moment of significantly above $0.1\,\mu_B$ may be reached.
Note that the maximum local moment in a collinear stripe structure is larger than
the one inferred from neutrons which averages over the oscillation period.

Taken together, we consider values of $V_s \gtrsim 0.07$ to be appropriate
to cause a negative Nernst signal.
From the experimental results reported in Ref.~\onlinecite{daou2009} and our results
in Fig.~\ref{ybconernst}, we then would infer that field strengths of around
20--30\,T are sufficient to produce a large negative normal-state Nernst coefficient in
underdoped \ybco. We assume that effects of Landau quantization are negligible in this
regime.

\subsection{Period-10 stripe order}

\begin{figure}
\includegraphics[width=3.7cm]{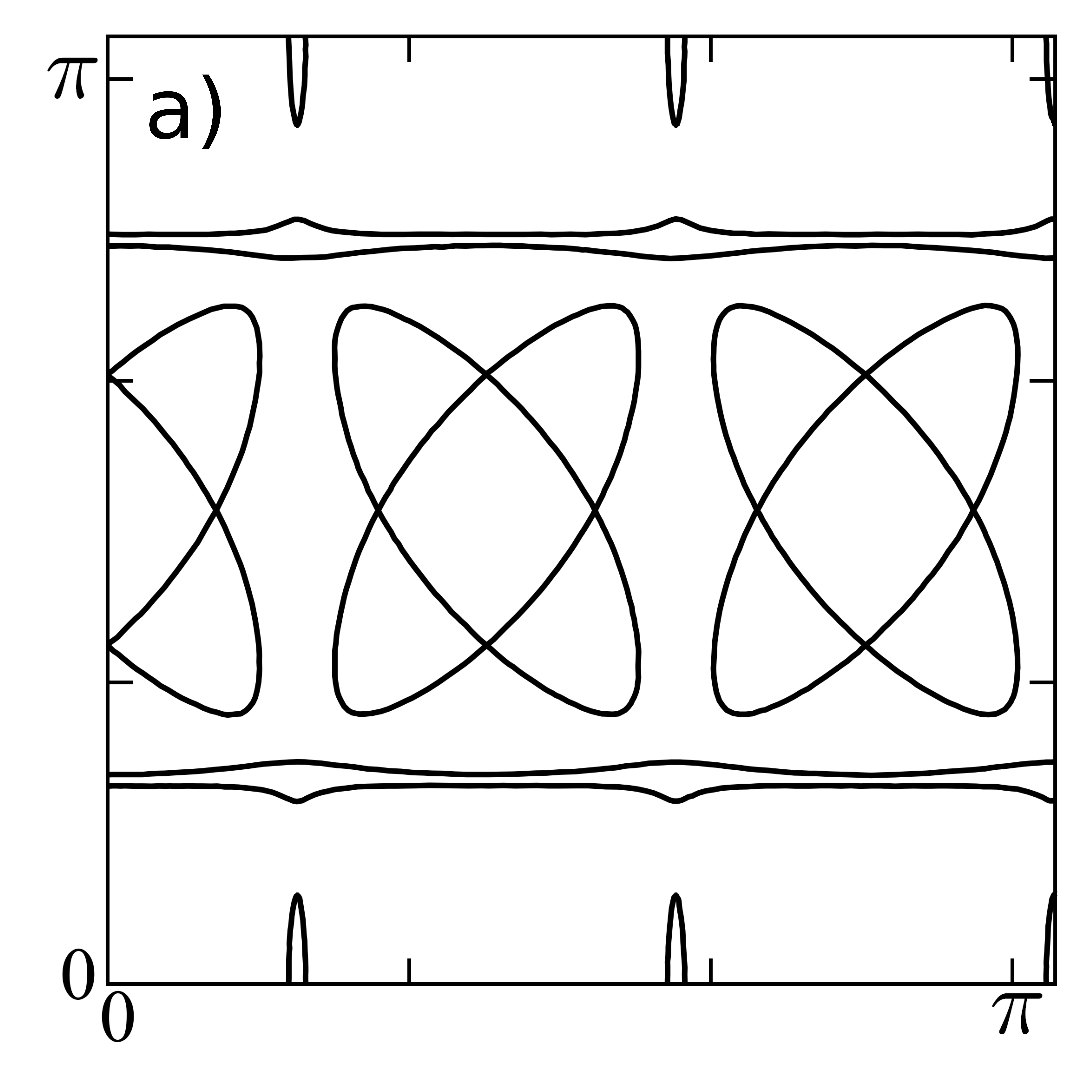} \includegraphics[width=3.7cm]{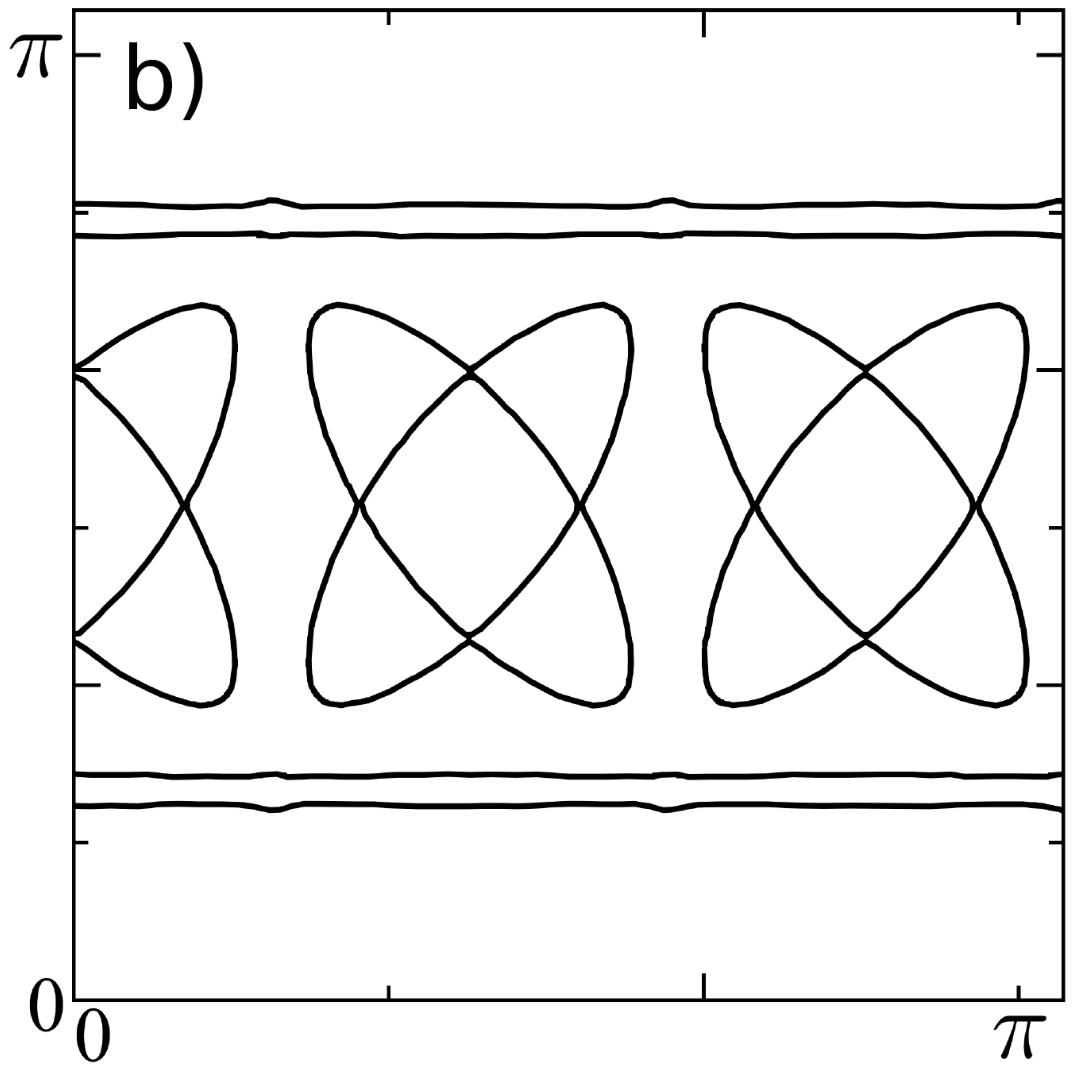}\\
\caption{
\label{underdopedfermi}
Fermi surfaces as resulting from period-10 stripe order,
plotted in the
first quadrant of the Brillouin zone of the underlying square lattice.
Pure spin stripe order with $V_s=0.08$\,eV produces both electron-like and hole-like closed orbits, see a).
Adding additional charge stripe order with $V_c=0.07$\,eV eliminates the electron like orbits and the remaining
closed orbits are all hole-like, see b).}
\end{figure}

Assuming $\epsilon_s=x$ for doping $x<1/8$ (as observed in 214 cuprates), a doping of $x=0.1$
corresponds to ${\bf Q}_s=\pi(4/5,1)$, leading to period-10 spin stripe order.
In this case, it is not possible that both charge and spin
modulations have extrema positioned on the bond centers, and we will assume a site-centered
stripe geometry in the following.
\begin{figure}
\includegraphics[width=7.0cm]{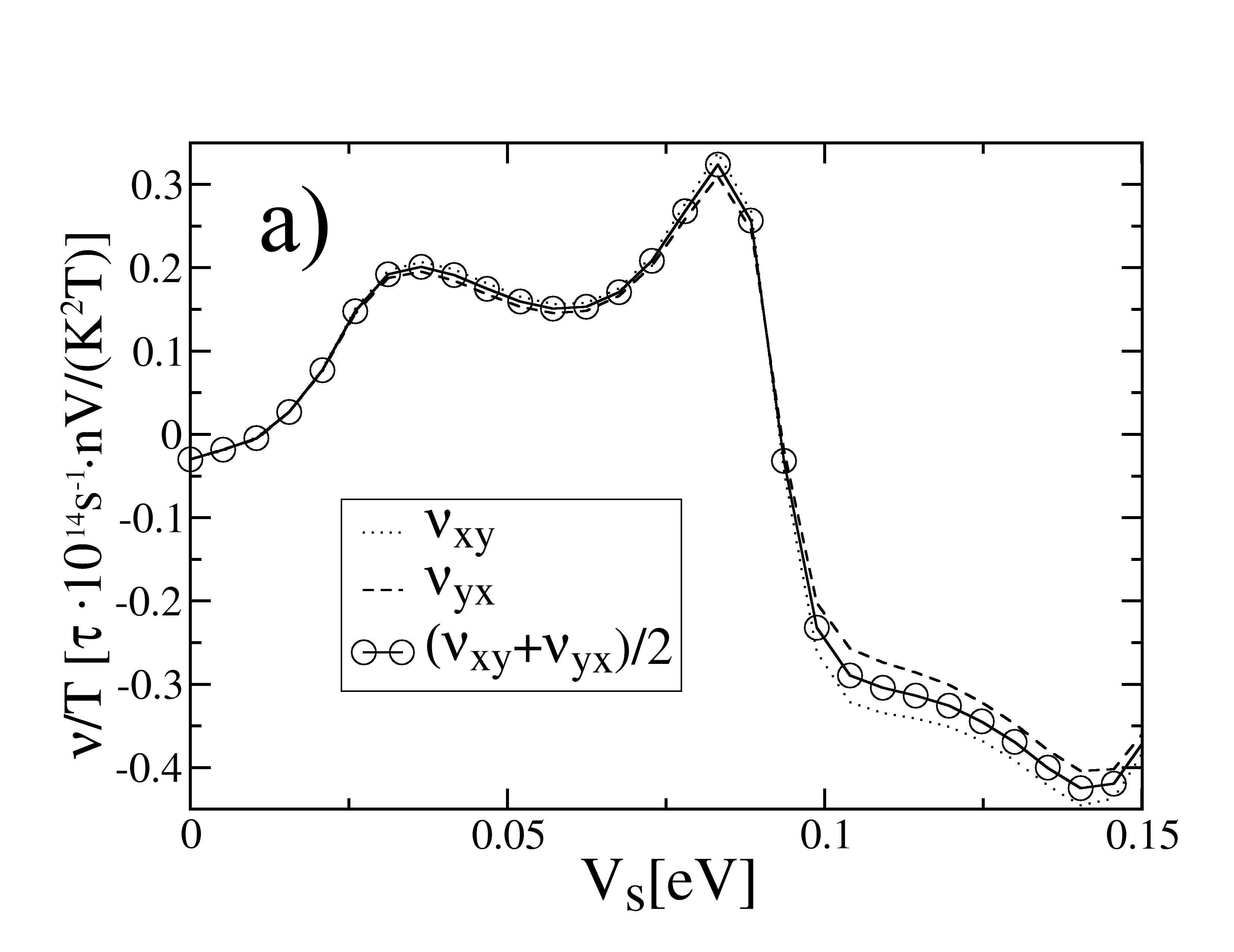}\\
\includegraphics[width=7.0cm]{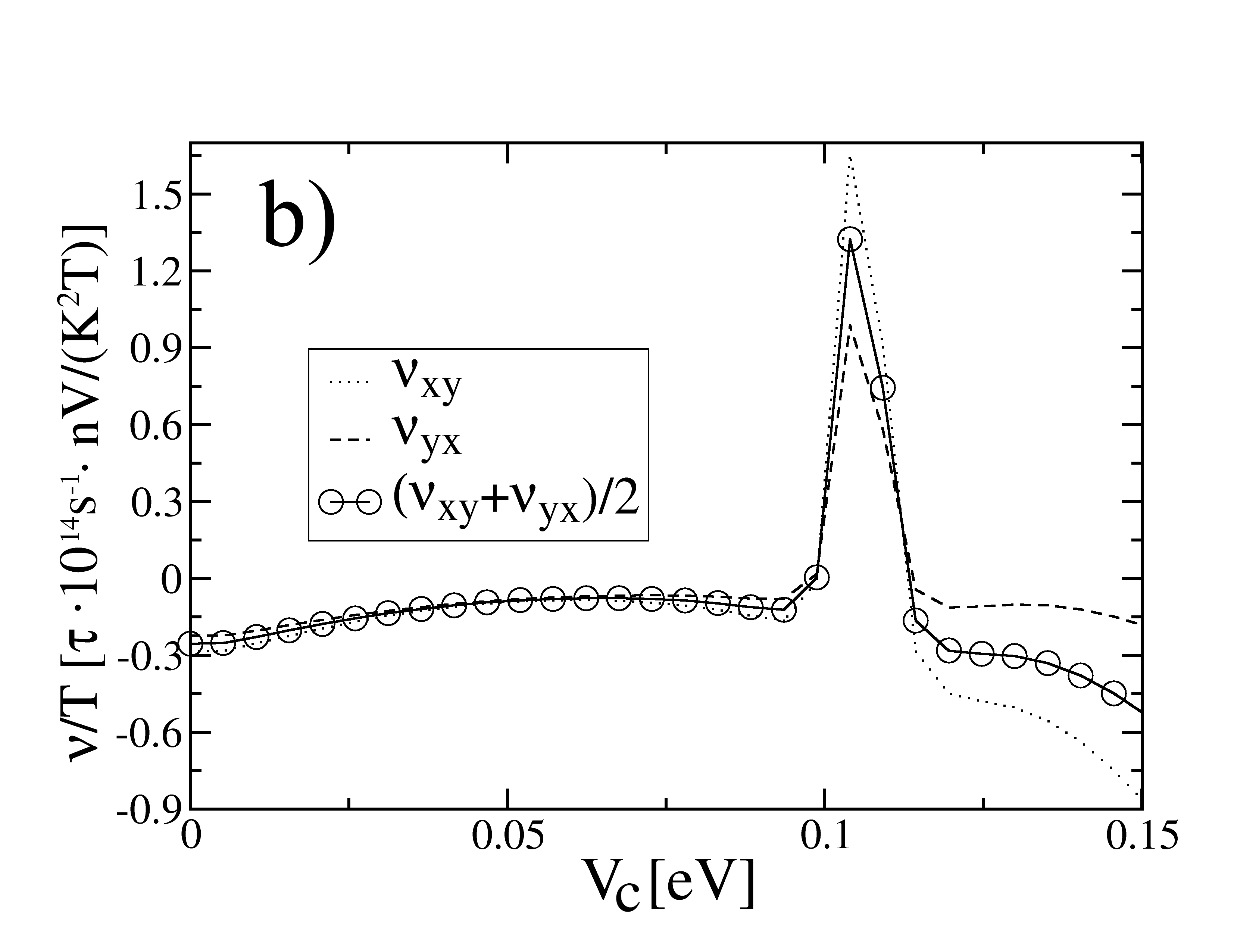}\\
\caption{\label{underdopednernst} Nernst effect for site-centered period-10 stripe order. Spin only stripe
order (a) ) leads to a negative Nernst coefficient for spin potentials above $V_s=0.09$\,eV. Adding additional
charge order to a spin potential of $V_s=0.1$\,eV  does not change the sign of the Nernst
coefficient for charge potentials $V_c\leq 0.1$\,eV, which correspond to realistic charge modulations of up to $30\%$ ( b)).}
\end{figure}
The Hamiltonian matrix corresponding to this type of order is thus analogous to the model
formulated in Eq.~(\ref{16bands}) and can be expressed by a $10\times 10$ matrix with the
appropriate ordering wavevector. We neglect a corrugation of the Fermi surface along the
z-direction, which has so far only been observed in YBCO. Typical Fermi surfaces
resulting from this model are described in Fig.~\ref{underdopedfermi}.

The Nernst coefficient resulting from pure spin stripe order shows a change to negative sign
at a spin potential strength of $V_s \simeq 0.09$\,eV corresponding to a maximal local moment of $0.25\mu_B$,
remaining negative up to a maximal ordered moment of $0.4\mu_B$,
see Fig.~\ref{underdopednernst} a). The negative sign can be
explained by the shrinkage of the small electron-like pockets shown in Fig.~\ref{underdopedfermi}
upon increasing $V_s$ above $V_s=0.1$\,eV, leading to a dominance of the closed hole-like orbits. As we
checked numerically, for these orbits $\alpha_{xx}\sigma_{xy} \gg \alpha_{xy} \sigma_{xx}$.
\begin{figure}
\includegraphics[width=8.0cm]{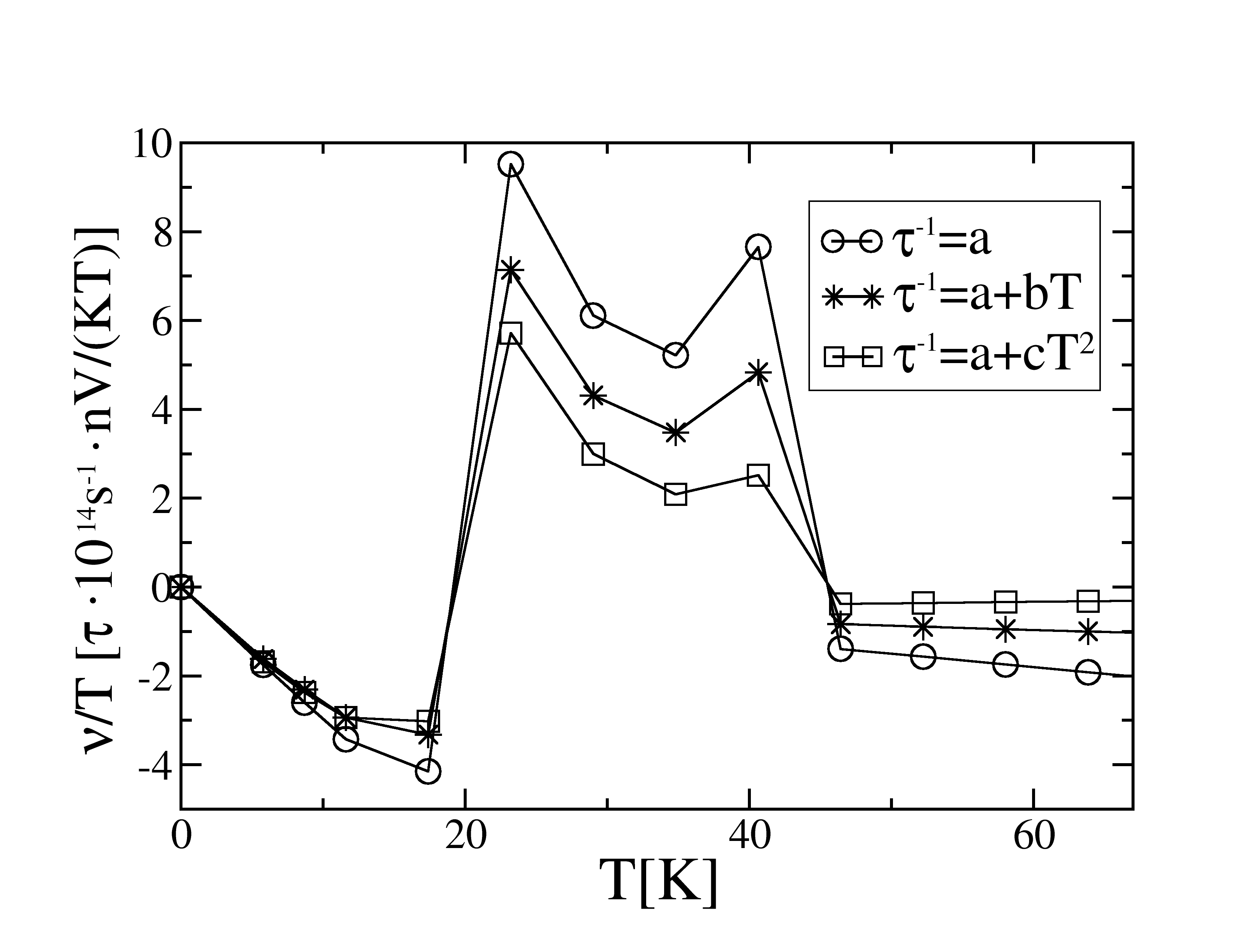}\\
\caption{
\label{10nernstthermal}
Nernst effect for period-10 stripe order at finite
temperatures. Upon decreasing temperature to below about 25\,K, the Nernst coefficient changes sign
and becomes negative. Upon increasing temperature above  about 25\,K, the coefficient
becomes positive and significantly enhanced. Slightly below the ordering temperature
$\Tsp\simeq50$\,K, the coefficient becomes negative again. The different scattering rates have been parameterized
with $a=\tau_0^{-1}$, $b=a/70$\,K and $c=a/800$\,K$^2$, and we set $V_0=0.12$\,eV.
}
\end{figure}
Since both $\alpha_{xx}$ and $\sigma_{xy}$ are positive for hole-like carriers,
the resulting Nernst coefficient has to be negative. Adding charge order has the effect
to finally eliminate the electron-like orbits (see Fig.~\ref{underdopedfermi} b).
This stabilizes a negative Nernst coefficient for charge potentials
corresponding to up to $30\%$ charge modulation, see Fig.~\ref{underdopednernst} b).

A discussion of the finite temperature properties of the Nernst coefficient is analogous to
the case of period eight stripe order.
Assuming a mean-field dependence $V_s(T)=V_0\sqrt{1-T/\Tsp}$ with $V_0=0.12$\,eV
and $\Tsp=50$\,K taken from neutron scattering experiments,\cite{ichikawa2000}
the resulting Nernst coefficient shows the two sign changes depicted in Fig.~\ref{10nernstthermal}.
These features are robust against specific parameterizations of the quasiparticle scattering
rate $\tau^{-1}$, as long as its temperature dependence is not too strong.
In conclusion, in underdoped \ndlsco\ samples with hole concentrations of about $x=0.1$ our result
predicts a negative peak in the Nernst coefficient as a function of temperature.
To observe this peak, eventually large magnetic fields have to be applied in order
to increase spin stripe order and to decrease vortex contributions to the Nernst coefficient.


\section{Influence of pseudogap and local pairing}
\label{pairing}

The model calculations presented so far have assumed the existence of metallic
quasiparticles, with a large Fermi surface in the underlying symmetry-unbroken state.
In underdoped cuprates, pseudogap phenomena are prominent in the temperature range
$\Tc < T < T^\ast$ where $T^\ast$ is the pseudogap temperature.
According to photoemission experiments on \bsco, \cite{kanigel06} the Fermi surface is partially gapped,
with Fermi arcs remaining near the Brillouin zone diagonals.
In stripe-ordered \lbco, \cite{he09,valla06} only nodal points appear to survive as low-energy excitations
below the stripe-ordering temperature.

Although many theories have been proposed to explain the pseudogap regime --
ranging from phase-fluctuating preformed Cooper pairs over competing orders to Mott
physics and strong short-range antiferromagnetic fluctuations --
its origin is still unclear.\cite{timusk,norman_rev,leermp}
As already mentioned in the introduction, experimental data suggests that
phase-fluctuating Cooper pairs alone cannot fully account for the observed pseudogap
phenomena.
With a lack of satisfactory descriptions of the pseudogap phase, we restrict ourselves to
a qualitative discussion in the following.

Regarding the relation between pseudogap and enhanced Nernst coefficient at intermediate
temperatures, different scenarios are conceivable, namely
pseudogap and Nernst coefficient may be caused by (i) the same or (ii)
different phenomena.
While the resistively defined pseudogap temperature seems to coincide
with the onset of a rapid change in the Nernst coefficient for dopings
above 1/8, the normal-state Nernst coefficient is distinctly peaked near
this doping, whereas the pseudogap continuously increases as the doping is
reduced.
We interpret this as evidence for scenario (ii).
Then, the effect of translational symmetry breaking on the Nernst coefficient may be
investigated, without fully accounting for (other) possible sources of
pseudogap phenomena -- this is the logic underlying the approach presented in this paper.
(Note there is little doubt that the experimentally seen
strong enhancement of the Nernst coefficient at temperatures near $\Tc$ is caused by
superconducting fluctuations.)

In strong magnetic fields and at low temperatures, it is conceivable that
the dominant source of corrections to the quasiparticle picture is given by
phase-fluctuating pairing,
with the phase incoherence becoming maximal near $H_{c2}$.
One possible explanation how phase fluctuations of the superconducting order parameter
are compatible with most of the phenomenology of the underdoped cuprates has been recently invoked
in Ref.~\onlinecite{senthil2009}.
In particular, in strong magnetic fields, appropriate to recent measurements of the Nernst and Hall effects,
the influence of phase fluctuations of the superconducting order parameter was argued to lead
mainly to a quasiparticle renormalization.
The scattering of the quasiparticles on a fluctuating $d$-wave order parameter
is described by the self-energy correction\cite{senthil2009}
\begin{equation}
\Sigma(\bk,\omega)= \Delta^2_{0 \bk} \frac{-i\omega+\varepsilon_\bk }{\omega^2+\varepsilon_{\bk}^2+\pi\Gamma^2}
\end{equation}
where $\Gamma$ is the phase decoherence rate of the order parameter amplitude $\Delta_0$
and $\Delta_{0 \bk}=\frac{\Delta_0}{2}(\cos(k_x)-\cos(k_y))$.
This correction leads to the renormalized quasiparticle dispersion $E(\bk)=\varepsilon(\bk)+\Sigma(\bk,\omega=0)$
\begin{equation}
E(\bk)=\epsilon_{\bk}\left(1+ \frac{\Delta^2_{0\bk} }{\varepsilon_{\bk}^2+\pi\Gamma^2}\right) \ .
\end{equation}
Remarkably, the original Fermi surface remains unchanged, and only renormalization of
band masses and quasiparticle velocities by a factor
$1+ \Delta^2_{0\bk} / (\varepsilon_{\bk}^2+\pi\Gamma^2)$
occurs. Since $\Gamma$ is of the order $\Delta_0^{-1}$
near $H_{c2}$, we may expect no qualitative change of transport properties
due to phase incoherent pairing at magnetic fields of the order $H_{c2}$.

A final remark on Fermi surfaces: While various photoemission experiments suggest truncated
Fermi surfaces in the form of arcs in the pseudogap regime, other experiments allow for
an interpretation in terms of Fermi pockets, which may be the result of symmetry-broken
states (as, in our case, stripes).
The issue of arcs vs. pockets is not settled, however, it has been proposed that both
matrix-element effects and disorder are responsible for the invisibility to photoemission
of parts of the pockets.\cite{sudip03,imada08}


\section{Summary}
\label{experiments}

We have calculated the normal-state Nernst coefficient in cuprates
in the presence of stripe-like translational symmetry breaking. The calculations were
based on a simple quasiparticle picture, combined with a Boltzmann equation approach.
The results demonstrate the role of Fermi pockets for a large quasiparticle Nernst coefficient.
The existence of such pockets depends on details of the symmetry breaking order;
in particular charge order alone does not easily generate pockets, but spin order is required.
Depending on both spatial period and amplitude of the stripe order, both positive and
negative Nernst coefficients can be generated, with sign changes as function of the stripe
amplitude which can be traced back to topological changes of the Fermi surface.

A robust positive Nernst signal was mainly found for period-8 modulated antiferromagnetic
order with ${\bf Q}_s=\pi(3/4,1)$,
appropriate for cuprates with doping levels $x\geq1/8$, as long as the magnetic
order is not assumed to be unrealistically strong.
For small charge modulation, there is little qualitative difference between bond-centered
and site-centered stripes; for larger charge modulation, site-centered stripes
tend to destroy hole-like orbits and induce a negative Nernst coefficient.
Charge order alone generates a small and typically negative Nernst coefficient.
Finally, open orbits contribute a small Nernst signal only, because off-diagonal
transport coefficients are small for quasi-one-dimensional bands.

The single-layer Nernst signal was naturally found to be anisotropic, but the anisotropy
was small, $\nu_{yx}$/$\nu_{xy} \in [0.5,2]$, for most parameter sets. Exceptions were states
with period-4 modulated charge order shown in Fig.~\ref{chargenernst}. In these cases,
hole pockets with large aspect ratios as shown in Fig.~\ref{purestripes}a)
can be present in the Fermi surface and can lead to large anisotropies of the Nernst signal.

\subsection{Relation to experiments}

Let us connect these results to experimental ones for the Nernst coefficient in cuprates.
Clearly, both quasiparticles and phase-fluctuating pairing will contribute to the
Nernst coefficient, with the latter not being part of the calculation presented in this
paper. This pairing-induced piece of the Nernst coefficient has been studied before
\cite{huse02,ussi03,podol07,raghu08} and is believed to dominate in a temperature region
near the superconducting $\Tc$, whereas an extra piece has been identified at elevated
temperatures in \ndlsco.\cite{taill09}
Our positive quasiparticle Nernst coefficient for period-8 stripes,
Figs.~\ref{nernstdoping} and~\ref{temperaturenernst},
is in qualitative agreement with these experimental results.
As function of temperature, the quasiparticle Nernst signal peaks below the charge
ordering temperature $\Tch$, vanishes linearly as $T\to0$ and becomes negative at high $T$,
Fig.~\ref{temperaturenernst}. (Adding a pairing-induced positive peak at low $T$ would
give a temperature dependence similar to experiment.)
Experimentally, the temperature maximum of the extra piece in the Nernst signal appears
to be above $\Tch$, which may be explained in terms of strong precursor stripe
fluctuations not captured in our mean-field theory.
The doping dependence of the quasiparticle Nernst signal in the doping
range $0.12<x<0.24$, Fig.~\ref{nernstdoping}, is in qualitative agreement with experiment
as well.

For magnetic modulation periods larger than 8 sites, the quasiparticle Nernst signal
displays sign changes as function of the modulation amplitude. From this, we predict sign
changes in the Nernst signal as function of temperature (in compounds with
well-established stripe order) or as function of applied field (if the order is primarily
field-induced).
Indeed, in a recent experiment\cite{daou2009} on \ybco\ at $y=6.67$, corresponding to a doping
level of 0.12, the Nernst effect at a field of 28\,T was found to be negative in the
low-temperature limit.
The signal showed substantial field dependence for smaller fields, with large positive
contributions near $\Tc$ due to superconducting fluctuations, but those have been argued to be
negligible in the regime above 25\,T.
Assuming that such fields induce sizeable SDW order with a modulation period larger than 8
(note that the observed spin correlations\cite{hinkov08} in \ybco\ do not follow the relation
$\epsilon_s\simeq x$), these findings could be consistent with our calculations.
Clearly, experiments on more underdoped \ybco\ samples are called for.

The huge anisotropy of the Nernst signal, found recently in \ybco\ at intermediate
temperatures,\cite{taill10b} cannot be easily explained in terms of magnetic stripe
states.
Instead, an interpretation\cite{hackl10} in terms of nematic order near a van-Hove
singularity appears more appropriate, while stripe order may set in at lower temperatures
(where indeed the experimental Nernst anisotropy decreases).


\begin{acknowledgments}
We acknowledge useful discussions with L.~Fritz, L.~Taillefer, A.~Wollny, and J.~Zaanen.
We also thank A. Millis and M. Norman for clarifications on Ref.~\onlinecite{millis07}.
A.H. and M.V. were supported by the DFG through the SFB 608 (K\"oln)
and the Research Units FG 538 and FG 960. S.S. was supported by the
U.S. National Science Foundation under grant DMR-0757145, by the FQXi
foundation, and by a MURI grant from AFOSR.
\end{acknowledgments}



\begin{thebibliography}{99}
\bibitem{impfoot}
One could go beyond the assumption of a constant $\tau_0$ by employing
a golden-rule calculation of the scattering rate from point-like static impurities.
As a result, $\tau$ would increase in the ordered state and would acquire some
energy dependence (due to gap opening).
However, as the quasiparticle Nernst coefficient appears to be dominated by
band-structure effects, we neglect this physics here.
\end{thebibliography}
\end{document}